\documentclass[final]{siamltex1213}

\usepackage{amssymb,amsmath,amsfonts,psfrag,overpic}
\newcommand{\figref}[1]{Figure~\ref{#1}}

\newcommand{\ud}{\mathrm{d}}
\newcommand{\bo}[1]{\boldsymbol{#1}} 

\title{Diffusion in spatially varying porous media} 

\author{Maria Bruna\footnotemark[2]\ \footnotemark[3] \and S. Jonathan Chapman\footnotemark[2]}

\begin{document}
\maketitle
\slugger{siap}{xxxx}{xx}{x}{x--x}

\footnotetext[2]{Mathematical Institute, University of Oxford, Radcliffe Observatory Quarter, Woodstock Road, Oxford OX2 6GG, United Kingdom (bruna@maths.ox.ac.uk, chapman@maths.ox.ac.uk).}  
\footnotetext[3]{Department of Computer Science, University of Oxford,
 Parks Road, Oxford OX1 3QD, United Kingdom. This author's work was supported by a Junior Research Fellowship of St John's College, Oxford, and by the EPSRC Cross-Discipline Interface Programme (grant number EP/I017909/1).}

\begin{abstract}
The problem of diffusion in a porous medium with a spatially varying porosity is considered. The particular microstructure analyzed comprises a collection of impenetrable spheres, though the methods developed are general. Two different approaches for calculating the effective diffusion coefficient as a function of the microstructure are presented. The first is a deterministic approach based on the method of multiple scales; the second is a stochastic approach for small volume fraction of spheres based on matched asymptotic expansions. We compare the two approaches, and we show good agreement between them in a number of example configurations.   
\end{abstract}



\pagestyle{myheadings}
\thispagestyle{plain}
\markboth{M. BRUNA AND S. J. CHAPMAN}{DIFFUSION IN SPATIALLY VARYING POROUS MEDIA}

\section{Introduction}
\label{sec:intro}

The  macroscopic transport of solutes in porous media depends
critically on microscopic features such as the structure of the porous matrix
 and the nature of the interactions between the solid and liquid
phases. 
On the other hand, 
the complexity of the microscopic problem means that in practice it is
often desirable to obtain a macroscopic effective-medium equation 
from which the  macroscopic transport can be
obtained directly.
This idea of upscaling is ubiquitous in many sciences. In particular, diffusive
transport in heterogeneous media occurs in hydrogeology (aquifers,
groundwater \cite{Fetter:2014tx,Freeze:1979vu}), contaminant transport
(water filtration with membranes), lithium-ion batteries \cite{Chung:2013kd}, and biological applications such as
porous biofilms \cite{Davit:2013fa} and intracellular transport
\cite{Novak:2009ck}.  

Efforts to find the form of the macroscopic  equation
and estimate the effective properties of the medium date back at least to the
19th century, when Maxwell gives an approximate expression for the
effective conductivity of a heterogeneous medium comprising small
(i.e.~well-separated) spheres of
one material distributed in another \cite[p.403]{Maxwell:1881vc}. 
Since then a great variety of approaches have been developed to obtain such
upscaled equations \cite{Cushman:2002fk}. A usual starting point is to
suppose that the solute undergoes a simple diffusion process in the
void or fluid phase $\Omega_v$ of the porous medium $\Omega$ so that the
evolution of the concentration of particles $C({\bf x},t)$ is
described by 
\begin{subequations}
\label{origin}
\begin{alignat}{2}
\label{origin_eq}
\frac{\partial C}{\partial t}  &= \nabla \cdot
\left ( D_0 \nabla C \right ),  \qquad &  {\bf x} &\in
\Omega_v,\\ 
\label{origin_bc}
0 &= {\bf n}  \cdot \left ( D_0 \nabla C \right ), &  {\bf x}
& \in \partial  \Omega_v, 
\end{alignat}
where $\partial \Omega_v$ is the solid-fluid interface and ${\bf n}$ is
the outward unit normal to $\Omega_v$.  The fraction of space
available to the diffusing species is the porosity $\Psi =
|\Omega_v|/|\Omega|$. If the solid phase is denoted by $\Omega_s$ (the
complement of $\Omega_v$) then the volume fraction of solid is $\Phi =
1-\Psi$. 
\end{subequations}
When the microstructure is fine we would like to be able to 
use an effective transport equation such as 
\begin{equation}
\label{ex_effective}
\frac{\partial c}{\partial t} = \nabla\cdot \left [ D_e
  \nabla c \right ], \qquad {\bf x} \in \Omega \equiv \Omega_v \cup \Omega_s, 
\end{equation}
where $D_e$ is an effective diffusion coefficient, which will depend
both on $D_0$ and the geometry of $\Omega_v$.
Here $c$ is the homogenized solute concentration (whose
definition will be made precise in the next section). Note that $c$ is defined
throughout the material, whereas $C$ is defined only in the fluid
region. 

In general, the homogenization procedure will depend on both the porous medium structure and the transport processes within the medium. A large number of homogenization techniques have been developed over the years \cite{Cushman:2002fk}. These can be divided into two very broad categories, deterministic and stochastic techniques. Deterministic techniques include techniques such as volume averaging \cite{whitaker:1999vj}, multiple scales \cite{SanchezPalencia:1980dg} and the generalized Taylor--Aris--Brenner method of moments \cite{Brenner:1980cq}. Here the averaging relies on the separation of scales between the microstructure and the macroscopic material, and is a local spatial average. Many techniques, included the method of multiple scales and the generalized Taylor--Aris--Brenner method of moments, assume some periodicity in the microstructure. In this case, the microscopic lengthscale measures variations within a period cell and the macroscopic lengthscale measures variations within the macroscopic region of interest. More specifically, the method of multiple scales constructs a family of problems involving the ratio $\delta$ between these two lengthscales and uses asymptotic expansions in $\delta$ to study systematically the convergence as $\delta \to 0$ to a limit problem. Since $\delta$ is assumed to be small, the resulting family of problems contains rapidly oscillating (periodic) coefficients. Thanks to its systematic nature, this method has formed the basis of a whole field in mathematics known as mathematical homogenisation \cite{Bensoussan:1978ti}.

\begin{figure}[bth]
\begin{center}
\begin{overpic}[width=.3\linewidth,tics=10]{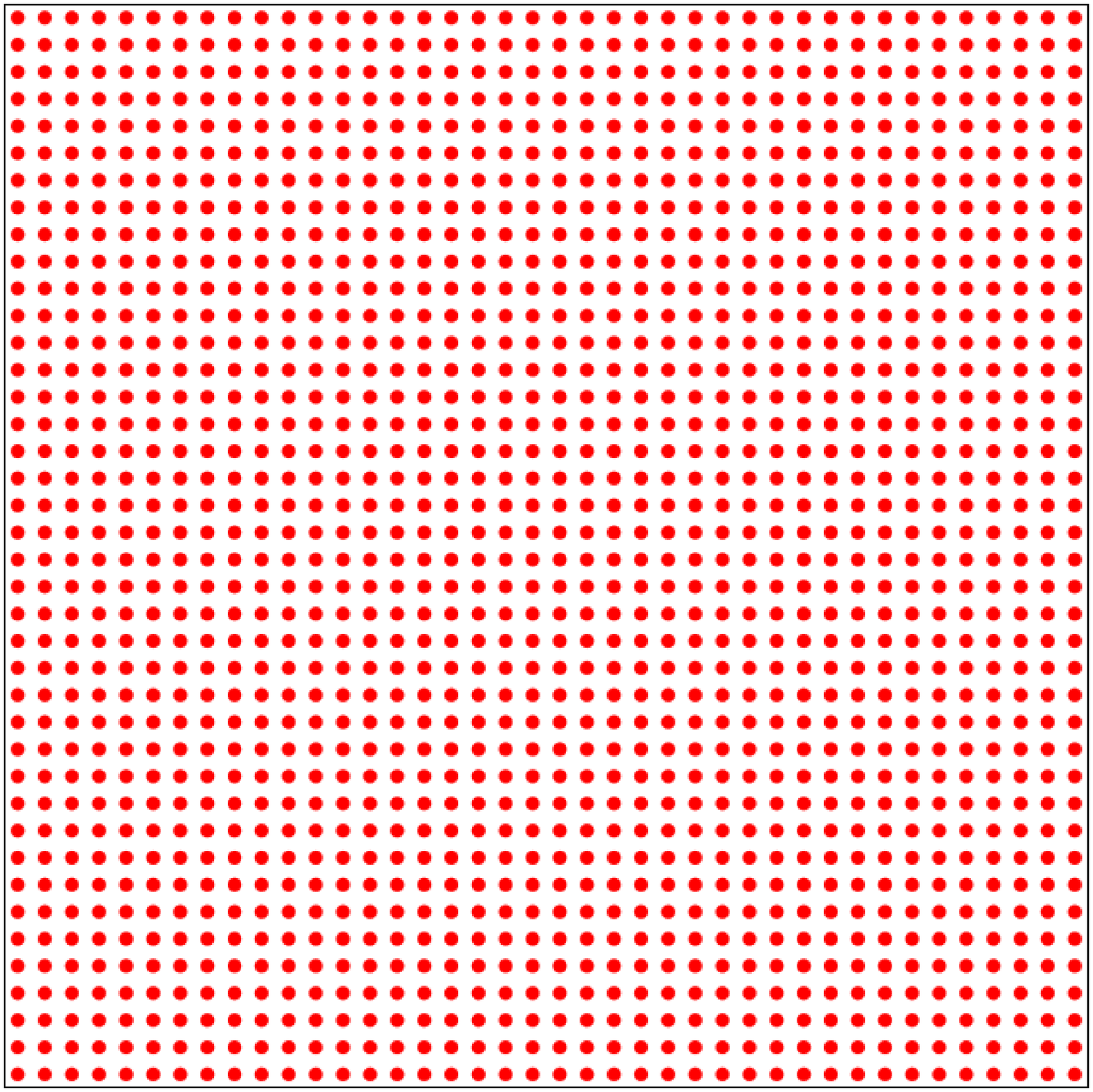} 
\put(-10,90){\small (a)}
\end{overpic}
\hspace{1cm} 
\begin{overpic}[width=.3\linewidth,tics=10]{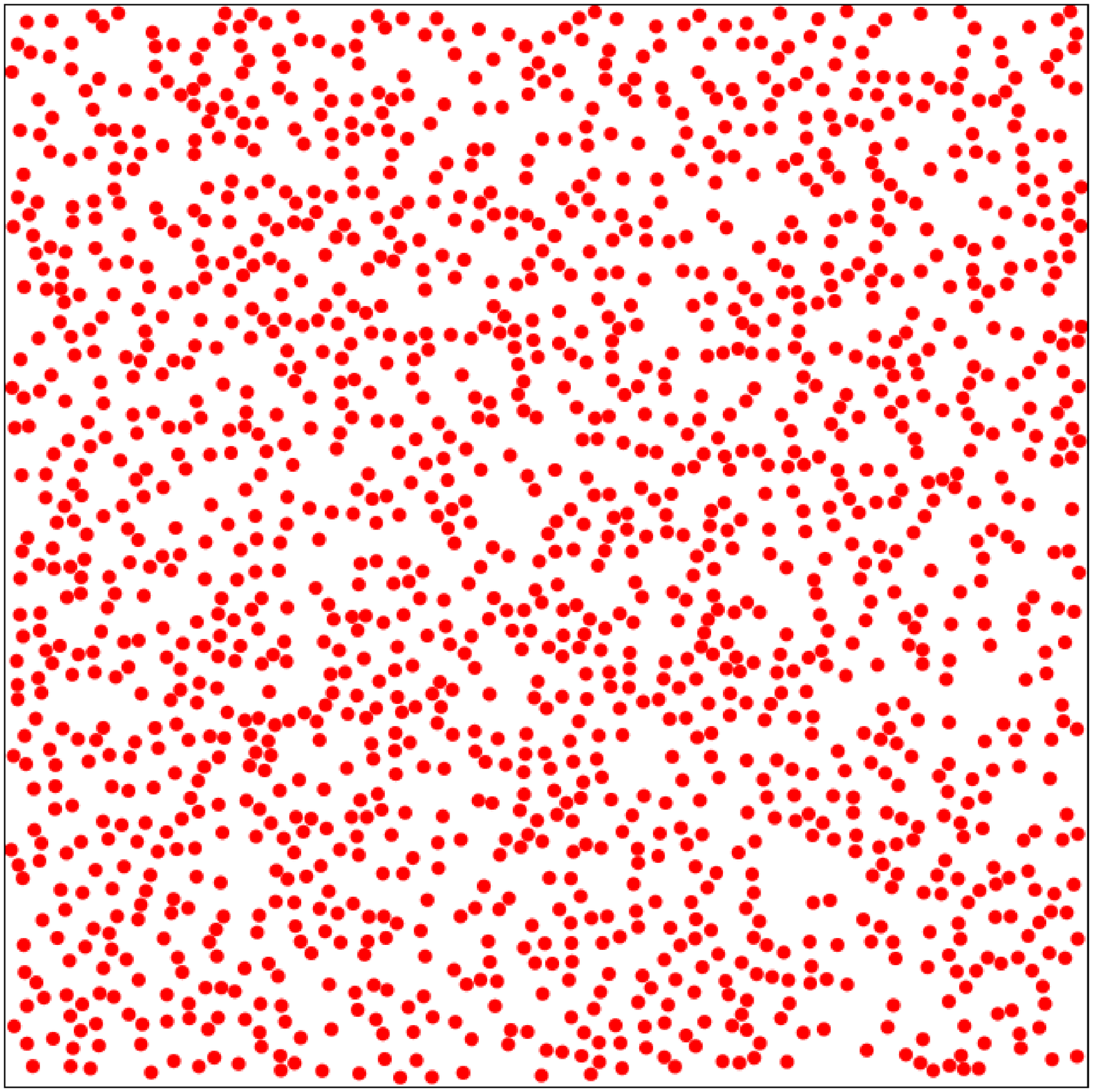} 
\put(-10,90){\small (b)}
\end{overpic}
\caption{Two examples of porous media. (a) Periodic
  structure (with spherical inclusions on a square lattice). (b) One sample from a random structure, with spherical inclusions
  uniformly distributed and with non-overlapping constraints. Both
  media correspond to a solid volume fraction of $\Phi = 20\%$.} 
\label{fig:det_ran}
\end{center}
\end{figure}

In terms of the example above, deterministic approaches upscale from \eqref{origin} to \eqref{ex_effective} by assuming a particular given geometry of the solid matrix $\Omega_s$ from which the effective coefficient $D_e$ can be computed. For example, it is common to consider either very simple periodic structures [such as the array of spheres depicted in \figref{fig:det_ran}(a)] or to consider a disordered unit cell  representative of the material as a whole, which is then extended periodically \cite{Torquato:1991hj}.
In the case of a periodic array of spherical inclusions, $D_e$ can be computed exactly using Rayleigh's multipole method \cite{Rayleigh:1892gq}, leading to an infinite system of
algebraic equations. More complex microscopic structures can be dealt
with via multiple scales (see for example \cite{Richardson:2011cb}) or volume
averaging \cite{whitaker:1999vj}. Although these two methods are
based on different underlying principles, they result in the same averaged equations \cite{Davit:2013dp}. These equations contain effective parameters (such as $D_e$) which depend on the microscale and are evaluated using the solution of a \emph{cell problem} in the
periodic unit cell (in multiple scales) or  the solution of a
closure problem in a 
\emph{representative elementary volume} (in volume averaging).  
For a review of the application of the method of volume averaging
in ordered and disordered porous media we refer the reader to
\cite{Quintard:1993bl}.  

The assumption of periodicity can be seen as artificial or too idealistic when modelling real heterogeneous media. As a result, an obvious progression came in the form of homogenization techniques for random media, in an attempt to reflect the uncertainty caused by the high degree of heterogeneity as well as the lack of experimental data \cite{Dagan:1989gz}. In its simplest setting, this can be seen as replacing the periodicity assumption in the multiple scales method by stochastic periodicity (or statistical homogeneity) \cite{Kozlov:1985ge,Papanicolaou:1979tm}. In particular, this means that the problem now contains random coefficients rather than periodic ones, and the homogenization consists of taking an averaging window of size $\delta$ large enough so that ergodicity holds, that is, that a spatial average is equivalent to an ensemble average. A rigorous derivation of the limit problem as $\delta \to 0$ is quite challenging in general, but has been done in specific cases such as steady heat conduction \cite{Kozlov:1985ge,Papanicolaou:1979tm} or, in a discrete setting, in a random conductance model for a network of resistors \cite{Gloria:2015bg}. In natural porous media such as soils and aquifers, the problem to be upscaled is usually more challenging than the cases above, as often one must consider the coupled problems for groundwater flow and solute transport. Stochastic approaches developed in the context of hydrology include the works of Cushman \cite{Cushman:1997wi}, Dagan \cite{Dagan:1989gz}, and Gelhar \cite{Gelhar:1993tx}. A common starting point is to take the hydraulic conductivity to be a random field, for example assuming it has a lognormal probability density function \cite{Dagan:1989gz} (p.~164). In particular, this implies that the Stokes velocity and the solute concentration are random fields as well. The classical approach is to linearize both the flow and the transport equations and to then take ensemble averages of the resulting problem \cite{Cushman:2002fk}.

In terms of the particular problem \eqref{origin}, a natural stochastic approach is to suppose that the microstructure is random and statistically homogeneous, with the effective properties arising by taking an ensemble average over different realizations of the microstructure \cite{Torquato:1991hj}. For example, one could assume that the
solid matrix $\Omega_s$ is composed of spherical inclusions uniformly distributed in $\Omega$ with a non-overlapping constraint [see \figref{fig:det_ran}(b) for one sample of such
configuration]. However, it is usually very difficult to estimate the
effective properties from such a description, since doing so requires
an infinite set of statistical correlation functions, which are
generally never known \cite{Torquato:1991hj}.  
Brown \cite{Brown:1955ft} provided a series form of the
effective constant  $D_e$ (for dielectric constants rather than
diffusivities, but the problem is analogous) and showed that 
the first correction term to $D_0$ depends only on the
solid volume fraction $\Phi$, but that all other terms require 
knowledge of the statistical distribution of the solid matrix. 
Thus  the best estimate one can obtain 
knowing only $\Phi$ and $D_0$ is equivalent to Maxwell's formula, namely
$D_e= D_0/(1 + \Phi/2)$.\footnote{The
  effective diffusivity reported was in fact $\tilde D_e = D_e
  (1-\Phi)$ in the current notation. Thus, the expression reported in
  \cite{Weissberg:1963vv} for example is $\tilde D_e/D_0 = (1-\Phi)/(1
  + \Phi/2)$. The two definitions result in the same macroscopic
  equation for homogeneous media, but not when the volume fraction varies on
  the macroscale, as we will see later.}  As a result, subsequent work
by Hashin and Shtrikman \cite{Hashin:1962ex} focused on using
variational methods to obtain
lower and upper bounds on $D_e$ as a function of $\Phi$, independent
of the statistics of the medium. Prager \cite{Prager:1963wa} developed a
method to introduce  two- and
three-particle correlation functions into such bounds. A specific
application of these 
variational principles in the case of a porous medium formed by
uniformly distributed spheres without non-overlapping constraints was
used by Weissberg \cite{Weissberg:1963vv}, who obtained the
 upper bound $D_e \leq D_0/(1-\frac{1}{2}\ln (1-\Phi
))$.\footnote{Again, this expression is reported as a volume averaged
  diffusivity, $\tilde D_e = D_e (1-\Phi)$ in
  \cite{Weissberg:1963vv}.} 
 
\def \scc {0.6}
\def \scl {1}

It is not completely clear which of these approaches is preferable in
any given situation, or how they compare to each other. As
pointed out in \cite{Lux:2010gd}, enforced periodic structures
 can have an important influence on the homogenized model in 
 low-porosity systems. On the other hand, the variational approaches
 used by Brown 
\cite{Brown:1955ft} or  Weissberg \cite{Weissberg:1963vv} are very
general and do not require any constraints on scale or periodicity.
They would be applicable also for so-called non-homogenizable media,
in which there is no clear separation of scales between micro and
macrostructures.

One situation in which all the approaches discussed above need to be
rethought is that in which there is macroscopic inhomogeneity in the
material as well as microscopic inhomogeneity. Such a situation would
arise in a material in which there is a non-uniform porosity on the
macroscale. It might be thought that since the porosity is varying
slowly it is locally constant, and the standard formulae can be used
albeit with a cell problem (and hence diffusivity) which
varies on the slow scale. However, we will see that this is not the case.

Despite its
acknowledged importance in many physical systems (for example, in the
design of membranes for water filtration, in groundwater flow systems
\cite{Fetter:2014tx}, and in Portland cement \cite{Rimmele:2008jh}), 
non-uniform porosity has rarely been accounted for in models and the
effect on diffusion is not well understood. One exception is
\cite{ValdesParada:2011dr}, which uses the method 
of volume averaging in systems with gradients in porosity.
Worryingly, though, they find that the effective diffusivity,
the porosity, and its gradient are highly dependent
upon the location of the centroid of the representative elementary volume.

In this paper we derive the effective transport equation for diffusion
in a porous
medium with gradients in porosity. We consider two types of porous
media, namely an ordered medium with quasi-periodic inclusions
and a disordered medium composed of randomly (but not uniformly) distributed
inclusions. The homogenization procedures for each of these 
media are fundamentally different. Thus, the purpose of the paper is
twofold: first, to study how macroscopic changes in the microstructure
affect the homogenized equation and, second, to compare
the two different approaches used to model diffusion in porous media.

We consider a simple diffusion process in the void or liquid phase,
where the solute has a constant diffusion coefficient $D_0$. The solid
phase is composed of spherical and non-overlapping inclusions, which
are impenetrable by the solute (that is, the diffusion coefficient is
zero in the solid phase). For most of this paper we consider a fixed
solid matrix (although we later remove
this assumption in the stochastic approach to consider 
moving obstructions).  

We obtain the effective diffusion coefficient as a function of the
microstructure of the medium,  and in particular the porosity. We find
that a non-uniform porosity on 
the macroscale results in an advection term in the homogenized
equation, as found by \cite{ValdesParada:2011dr}. This term accounts
for the biasing of the motion of solute particles towards regions of high
porosity. We also show that our two approaches, although very
different, can be reconciled to give
the same  macroscopic equation in the limit of small solid volume fraction.  

The article is organized as follows. In \S\ref{sec:determ} we consider a deterministic approach and we apply the method
of multiple scales to derive the effective transport equation in ordered porous media. We show how the method can be
adapted to non-uniform porosity and non-periodic structures, providing
they are locally periodic. We then consider the same problem for
stochastic porous media (disordered and non-periodic) in \S\ref{sec:stoch},
and derive an equivalent effective equation from the Fokker--Planck
description of the microstructure. In \S\ref{sec:comparision} we present two
numerical studies comparing the two approaches, both with each other
and with the exact solution of the corresponding microscopic problems. 
Finally, in \S\ref{sec:discussion}, we present our conclusions.

\section{An ordered porous medium}
\label{sec:determ}
We rescale length and time so that the rescaled domain has volume one, $|\Omega| = 1$ and the molecular diffusion coefficient is $D_0 =
1$.\footnote{We abuse the notation by  using the same symbols $\Omega$ and $D_0$ as in the dimensional problem \eqref{origin}.} We also rescale concentration so that $\int_{\Omega_v} C \ud {\bf x} = 1$. For simplicity we assume that $\Omega = [-1/2,1/2]^d$, where the
dimension $d$ may be $2$ or $3$.  
We suppose that the solid phase consists of $N_s$ spherical and
non-overlapping obstacles with radii $\epsilon_i$ and centers ${\bf r}_i$
for $i=1,\dots N_s$. We suppose that the radius is a slowly varying
function of position given by $\epsilon({\bf x})$, so that we may
write $\epsilon_i = \epsilon({\bf r}_i)$. Thus 
$\Omega_s = 
\cup_{i=1}^{N_s} B_{\epsilon({\bf r}_i)} ({\bf r}_i)$, where
$B_\epsilon({\bf r})$ is the $d$-dimensional ball of radius $\epsilon$
centered at ${\bf r}$. 

\begin{figure}[htb]
\begin{center}
\input{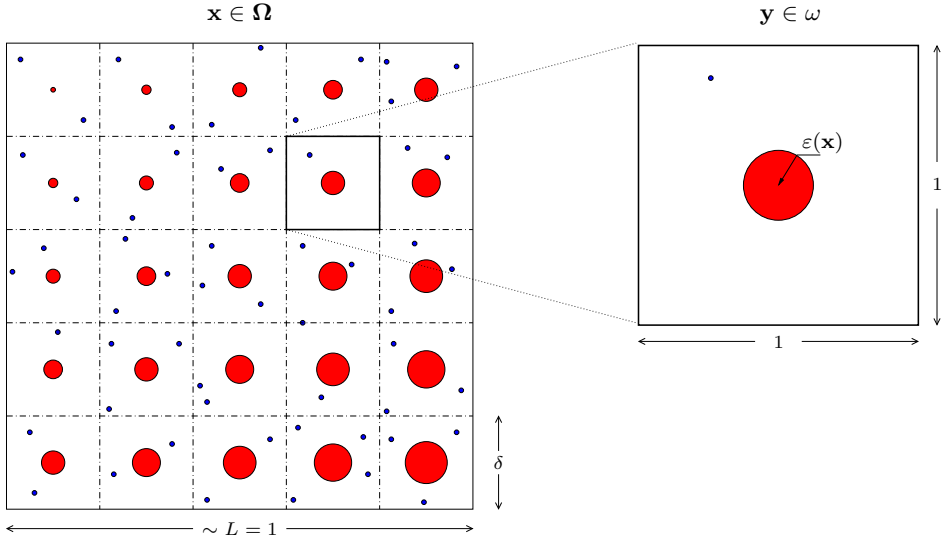}
\caption{Illustration of the problem geometry in two dimensions. Left:
  macroscale domain $\Omega$, with obstacles of radius $\epsilon({\bf
    x})$ placed in a periodic array. Right: microscale domain
  $\omega$, unit cell with an obstacle of radius $\varepsilon({\bf x})
  = \epsilon({\bf x})/\delta$. } 
\label{fig:periodicdomain}
\end{center}
\end{figure}

We begin by supposing that the 
centers ${\bf r}_i$ lie on a  regular square lattice with
period $\delta \ll 1$, as depicted in \figref{fig:periodicdomain}. 
Then
\eqref{origin} reduces to  
\begin{subequations}
\label{determin0}
\begin{alignat}{2}
\label{eq0}
\frac{\partial C}{\partial t}  &= \nabla^2  C ,
\qquad &  &{\bf x} \in \Omega_v,\\ 
\label{bc0}
 {\bf n}  \cdot   \nabla C&=0 , & & \mbox{ on }\| {\bf x} -
 {\bf r}_i\| = 
\epsilon({\bf r}_i), \quad 1\le i \le N_s, 
\end{alignat}
with $N_s = \delta^{-d}$. 
\end{subequations}
The local porosity  is then $\psi({\bf x}) = 1-
\delta^{-d} |B_{\epsilon({\bf x})}|$. The global porosity is the average
of the local porosities and is determined by the volume of all $N_s$
obstructions, namely 
\[
\Psi= \frac{1}{N_s} \sum_{i=1}^{N_s} \psi({\bf r}_i)  = 1 -
\sum_{i=1}^{N_s} |B_{\epsilon({\bf r}_i)}|. 
\]

\subsection{Asymptotic homogenization via the method of multiple scales}
\label{sec:multiplescales}

We use the method of multiple scales to derive an averaged (or
homogenized) model for the concentration $C$, valid over many obstacles, in
the limit $\delta \ll 1$. We retain ${\bf x}$ as the macroscale
variable, measuring distance on the scale of the whole sample, and
introduce ${\bf y} = {\bf x}/\delta$ as the microscale variable,
which measures distance over the scale of the obstacle
separation. 
As is usual in the method of multiple scales we treat these two
variables as independent. The extra freedom this gives is removed by
enforcing that the solution is exactly periodic in ${\bf y}$; small
variations from one unit cell to the next are thereby  captured through the
macroscale variable ${\bf x}$.

After introducing these two scales and using the chain rule spatial
derivatives in 
\eqref{determin0} transform 
according to  
\begin{equation}
\label{changevars}
\nabla  \longrightarrow   \nabla_{\bf x}  + \frac{1}{\delta} \nabla_{\bf y}.
\end{equation}
We introduce $\varepsilon({\bf x}) = \epsilon({\bf x})/\delta$ as the
the obstacle radius relative to the dimension of the unit cell. As
stated above, a crucial assumption we make is that this varies slowly,
that is, it depends on ${\bf x}$ but not on ${\bf y}$.
We denote the unit cell by ${\bf y} \in \omega \equiv [-1/2,1/2]^d$
and the solid phase by ${\bf y} \in \omega_s({\bf x}) = B_{\varepsilon
  ({\bf x})} ({\bf  0})$, with the fluid phase then given by ${\bf y}
\in \omega_v({\bf x}) = \omega\setminus\omega_s({\bf x})$ (see
right-hand side of \figref{fig:periodicdomain}).

The expression \eqref{changevars} allows us to transform \eqref{eq0}
according to the multiple-scales approximation. What is not clear is how we should write
\eqref{bc0} in multiple-scales form, given that the radius of the
obstacle depends on ${\bf x}$, so that the unit normal depends on
${\bf x}$ as well as ${\bf y}$. At first sight it would seem that
geometry dictates that 
\begin{equation}
{\bf n} = {\bf n}_{\bf y} \equiv -\frac{{\bf y}}{\|{\bf y}\|}\label{ny}
\end{equation}
and \eqref{bc0} should be written
\begin{equation}
\left(\frac{1}{\delta}\nabla_{{\bf y}}+ \nabla_{{\bf x}} \right) C
\cdot {\bf n}_{\bf y} 
= 0  \quad \mbox{ for }{\bf y} \in \partial 
\omega_v({\bf x}).  
\end{equation}
However, this is incorrect as it neglects the variation of ${\bf n}$
with ${\bf x}$.

The systematic way to derive the multiple-scales equivalent of
\eqref{bc0} is to introduce the function
\begin{equation}
\label{boundary}
\chi({\bf x}, {\bf y}) = \varepsilon({\bf x}) - \| {\bf y} \|,
\end{equation} 
for which the fluid--solid interface is the level set $\chi({\bf
  x}, {\bf y}) = 0$. 
Note that this idea can be extended to more complicated
interfaces in a straightforward manner. 
The normal is in the direction $\nabla\chi$, which we can now readily
put in multiple-scales form using \eqref{changevars}:
\begin{equation}
\label{normal_micro}
{\bf n} \propto \nabla_{\bf x} \chi + \frac{1}{\delta} \nabla_{\bf y} \chi = \nabla_{\bf x} \varepsilon ({\bf x}) - \frac{1}{\delta} \frac{\bf y}{\| \bf y\|}.
\end{equation}
This expression for the normal in multiple-scales form is
far from obvious, and is a crucial part in extending the theory in
\cite{Richardson:2011cb} to macroscopically varying cell
geometries. There are two contributions to the interface normal in
\eqref{normal_micro}; the first is due to the slow variation of the
obstacle radius, and the second is the  unit radial vector in the
microscale.

Thus writing \eqref{determin0} in multiple-scales form gives  
\begin{subequations}
\label{s1}
\begin{alignat}{2}
\label{eq1}
\delta^2 \frac{\partial C}{ \partial t}  &=
\nabla^2_{{\bf y}} C + \delta \nabla_{{\bf x}} \cdot \nabla_{{\bf y}}
C  + \delta \nabla_{{\bf y}} \cdot \nabla_{{\bf x}} C + \delta^2
\nabla^2_{{\bf x}} C & \quad    &{\bf y} \in \omega_v({\bf x}), \\ 
\label{bc1}
\nabla_{{\bf y}} C \cdot {\bf n}_{\bf y} & = -  \delta \nabla_{{\bf
    x}} C \cdot {\bf n}_{\bf y}  
- \delta \nabla_{{\bf y}} C \cdot \nabla_{{\bf x}} \varepsilon ({\bf
  x}) - \delta^2  \nabla_{{\bf x}} C \cdot \nabla_{{\bf x}}
\varepsilon ({\bf x})  
& &{\bf y} \in \partial \omega_v({\bf x}),  
\end{alignat}
\end{subequations}
where ${\bf n_y}$ (given in  \eqref{ny}) is the unit normal
vector into the obstacle in the microscale variable.

Our aim is to derive the equation for the averaged macroscopic
concentration $c({\bf x},t)$ over a representative  volume located at
position ${\bf x}$. With our periodic microstructure we may define this as
\begin{equation}
\label{average}
c({\bf x},t) = \frac{1}{|\omega|} \int_{\omega} C ({\bf x, y}, t) \,
\ud {\bf y} = \frac{1}{|\omega|} \int_{\omega_v({\bf x})} C ({\bf x,
  y}, t) \, \ud {\bf y}, 
\end{equation}
since $C\equiv 0$ in the solid phase. 
In the context of volume averaging, the unit cell $\omega$ is the
\emph{representative elementary volume} (REV), and is located at the
centre of each of the obstacles. The average $c$ is referred to as
\emph{volumetric} or \emph{superficial average}. It is related to  the
\emph{intrinsic 
  average}
\begin{equation}
\label{intrinsic_average}
\bar c({\bf x},t) = \frac{1}{|\omega_v ({\bf x})|}
\int_{\omega_v({\bf x})} C ({\bf x, y}, t) \, \ud {\bf y}, 
\end{equation}
through the porosity $\psi({\bf x})$: 
\begin{equation}
\label{relationPs}
c({\bf x},t) = \psi({\bf x}) \bar c({\bf x},t) \qquad \mbox{ where
} \quad  \psi({\bf
  x}) = \frac{|\omega_v ({\bf x})|}{|\omega|}. 
\end{equation}
With a spherical obstruction in the centre of each cell, the 
porosity is $\psi ({\bf x}) = 1 - |B_{ \varepsilon({\bf x}) }|$. 
The concentration $c$ is normalized in the whole space, whereas $\bar c$
is normalized in the available space.  

Following the standard multiple-scales method, we now seek a solution
to \eqref{s1} in the limit of small $\delta$ 
of the form  $C = C({\bf x},{\bf y}, t)$ which is periodic in ${\bf
y}$, while treating ${\bf x}$ and ${\bf y}$ as independent. 
Expanding in powers of $\delta$ as $C ({\bf x},{\bf y}, t)=
C^{(0)}( {\bf x}, 
{\bf y}, t) + \delta  C^{(1)}({\bf x},{\bf y}, t) + \delta^2  C^{(2)}(
{\bf x}, {\bf y}, t) + \cdots$ we find the
leading-order equations require that $C^{(0)}$ is
independent of  ${\bf y}$. 
At first order in $\delta$ we find
\begin{subequations}
\label{o1}
\begin{alignat}{2}
\label{eqo1}
\nabla^2_{{\bf y}} C^{(1)} &=  0 & \qquad   &{\bf y} \in \omega_v
({\bf x}), \\ 
\label{bco1}
\nabla_{{\bf y}} C^{(1)} \cdot {\bf n}_{\bf y} & = - \nabla_{{\bf x}}
C^{(0)} \cdot {\bf n}_{\bf y} & & {\bf y} \in \partial \omega_v({\bf
  x}),\\ 
\label{bcp1}
C^{(1)} & \quad  \textrm{periodic} & &\textrm{in} \ {\bf y}.
\end{alignat}
\end{subequations}
The solution of \eqref{o1} can be written as
\begin{equation}
\label{2:solo1}
C^{(1)}({\bf x}, {\bf y}, t) = -\nabla_{{\bf x}}  C^{(0)} ({\bf x}, t)
\cdot \bo \Gamma ({\bf x}, {\bf y}), 
\end{equation}
where $\bo \Gamma ({\bf x}, {\bf y})$ is a $d$-vectorial function,
whose components $\Gamma_i$  satisfy the following \emph{cell
  problem}: 
\begin{subequations}
\label{cellproblem}
\begin{alignat}{2}
\nabla^2_{{\bf y}} \Gamma_i &=  0 &\qquad   & {\bf y} \in \omega_v
({\bf x}), \\ 
\nabla_{{\bf y}} \Gamma_i \cdot {\bf n}_{\bf y} & =n_{y,i} & & {\bf y}
\in \partial \omega_v({\bf x}), \\ 
\Gamma_i & \quad  \textrm{periodic} & &\textrm{in} \ {\bf y},
\end{alignat}
\end{subequations}
where $n_{y,i}$ is the $i$th component of the unit vector ${\bf
  n}_{\bf y}$. We note that $\bo \Gamma$ varies with
the macroscale variable ${\bf x}$ because of the variation of
$\omega_v$ with ${\bf x}$. 

Proceeding to second order in the expansion of  \eqref{s1} leads to
the following problem for $C^{(2)}$: 
\begin{subequations}
\label{o2}
\begin{alignat}{2}
\label{eqo2}
\frac{\partial C^{(0)}}{ \partial t}  &= 
 \nabla_{{\bf x}} \cdot  \left (
\nabla_{{\bf y}} C^{(1)} +  \nabla_{\bf x} C^{(0)}  \right)   +
\nabla_{{\bf y}} \cdot \left( \nabla_{{\bf y}} C^{(2)} + \nabla_{{\bf
    x}} C^{(1)} \right) &\quad   &{\bf y} \in \omega_v ({\bf x}), \\ 
\label{bco2}
\nabla_{{\bf y}} C^{(2)} \cdot {\bf n}_{\bf y} & = -   \nabla_{{\bf
    x}} C^{(1)} \cdot {\bf n}_{\bf y} -   ( \nabla_{{\bf y}} C^{(1)} +
\nabla_{{\bf x}} C^{(0)}) \cdot \nabla_{{\bf x}} \varepsilon ({\bf x})  
& &{\bf y} \in \partial \omega_v({\bf x}), \\
\label{bcp2}
C^{(2)} &\quad \textrm{periodic in } \ {\bf y}. 
\end{alignat}
\end{subequations}
Integrating \eqref{eqo2} over  $\omega_v({\bf x})$ using the
divergence theorem, \eqref{bco2} and periodic boundary conditions on
$C^{(1)}$ and $C^{(2)}$ yields 
\begin{equation}
\label{solo3}
\begin{aligned}
\psi({\bf x})\frac{\partial C^{(0)}}{ \partial t}  =   & 
\int_{\omega_v ({\bf x}) } \! \! \nabla_{{\bf x}} \cdot \left (
\nabla_{{\bf y}} C^{(1)} +  \nabla_{\bf x} C^{(0)}  \right) \ud {\bf
  y} \\
  & - \int_{ \partial \omega_v({\bf x}) } \! \!  \left ( \nabla_{{\bf
    y}} C^{(1)} +  \nabla_{\bf x} C^{(0)}  \right) \cdot \nabla_{{\bf
    x}} \varepsilon ({\bf x}) \, \ud S_{\bf y}. 
    \end{aligned}
\end{equation}
Now using the transport theorem to switch the order of integration with
respect to ${\bf y}$ and differentiation with respect to ${\bf x}$ in the
term on the right-hand side  gives 
\begin{align}
\label{solo33}
\psi({\bf x}) \frac{\partial C^{(0)}}{ \partial t}   =
\nabla_{{\bf x}} \cdot   \int_{\omega_v ({\bf x}) }   \left (
\nabla_{{\bf y}} C^{(1)} +  \nabla_{\bf x} C^{(0)}  \right) \ud {\bf
  y}. 
\end{align}
Let  $J_{\bf \Gamma} ({\bf x}, {\bf y}) $ be the Jacobian matrix of
$\bf \Gamma$ given by  $(J_{\bf \Gamma})_{ij} = \partial \Gamma_i / \partial
y_j$. Then using \eqref{2:solo1}  we may write  
\begin{equation}
\label{matrix}
\nabla_{\bf y} C^{(1)} =  -  J_{\bf \Gamma}^T ({\bf x}, {\bf y})
\nabla_{\bf x} C^{(0)}. 
\end{equation}
Combining \eqref{solo33} and  \eqref{matrix} we arrive at the
following equation for the leading-order intrinsic average \mbox{$\bar c =
  C^{(0)}({\bf x},t)$}: 
\begin{subequations}
\label{sol_poreav}
\begin{align}
\label{solo4}
\psi({\bf x}) \frac{\partial \bar c}{ \partial t}   =
\nabla_{{\bf x}} \cdot  \left[ \psi({\bf x}) D_e({\bf x}) \nabla_{{\bf
      x}} \bar c \right], 
\end{align}
 where the effective diffusion tensor is given by
\begin{align}
\label{Dx} 
D_e({\bf x}) &= I_d - \frac{1}{\psi({\bf x})} \int_{\omega_v ({\bf x})
}    J_{\bf \Gamma}^T({\bf x}, {\bf y})  \, \ud {\bf y}, 
\end{align}
where  $I_d$ is the identity matrix of dimension $d$.  
\end{subequations}
The resulting model \eqref{sol_poreav} is equivalent to that derived by \cite{ValdesParada:2011dr} using a volume averaging approach.\footnote{See  their equations (16) and (17); in our notation, $\varepsilon_\gamma = \psi({\bf x})$, $\langle c_{A\gamma} \rangle^\gamma = \bar c ({\bf  x},t)$ and $  \bo D_\text{eff} = D_e({\bf x})$. However, we note that their cell problem (14) includes formally higher-order terms and hence is not exactly equal to our cell problem \eqref{cellproblem}.}

We see that if $\psi$ is independent of ${\bf x}$ then it may be
cancelled on both sides of \eqref{solo4} giving the usual homogenized
result from multiple scales. However, when the radius of the obstacles
varies with ${\bf x}$ the effect is felt not only through $D_e$ (due
to a cell problem that depends on ${\bf x}$) but also through the
fact that $\psi$ on the right-hand side appears {\em inside} the ${\bf
  x}$-derivative.  

\subsubsection{Comparison between averages: superficial  versus intrinsic}
\label{sec:porevolume}

The homogenized model \eqref{sol_poreav} for the intrinsic average
$\bar c$ is not yet an effective diffusion equation because of the
porosity $\psi$ multiplying the left-hand side of
\eqref{solo4}. Hence it is incorrect to refer to $\psi({\bf x})
D_e({\bf x})$ in \eqref{solo4} as an effective diffusion coefficient. 
However, if we rewrite it in terms of the volume average $c = \psi
\bar c$ we find
\begin{align}
\label{sol_inc}
\begin{aligned}
 \frac{\partial c}{ \partial t}   &=   \nabla_{\bf x} \cdot
 \left[ \psi  D_e   \nabla_{\bf x} \left(\frac{c}{\psi}\right) \right] =   \nabla_{\bf
   x} \cdot  \left[ D_e({\bf x})  \nabla_{\bf x} c -
   \frac{D_e({\bf x}) \nabla_{\bf x} \psi ({\bf x}) } { \psi ({\bf
       x}) } c  \right]. 
 \end{aligned}
\end{align}
Thus, we see that, when written in terms of $c$, the equation turns into a
classical advection--diffusion equation, with $D_e({\bf x})$ the
effective diffusion coefficient. The advection term \mbox{${\bf
  v} = D_e({\bf x}) \nabla_{\bf x} \psi ({\bf x}) / \psi ({\bf x})$}
arises when the porosity is non-uniform. We see that  solutes
diffuse with a bias towards 
regions of increased porosity.  

In the case of uniform porosity $\psi({\bf x}) \equiv \Psi$, the
intrinsic average $\bar c$ is simply a scaled version of the
volumetric average
$c$ and their respective  equations \eqref{sol_inc} and \eqref{solo4}
both reduce to a diffusion equation with constant diffusion coefficient
$D_e$ as in \eqref{ex_effective}.  However, when the porosity is
non-uniform the equations are different, and it is important to be
clear which average of $C$ we are dealing with.

\subsubsection{The diffusion tensor}

To evaluate the diffusion tensor $D_e({\bf x})$ in \eqref{Dx}, we must
solve the cell problem \eqref{cellproblem}. So far we have not used
that the obstacle in each cell is spherical, and in principle we could
solve the cell problem 
(numerically) for any obstacle shape. However, having a spherical
obstacle greatly simplifies the procedure. This is because, by
symmetry, we find that $D_e$ is a multiple of the identity tensor,  so
that we have a single scalar diffusion coefficient.  
 
We solve the cell problem \eqref{cellproblem} using COMSOL
Multiphysics and evaluate the integrals in \eqref{Dx}
numerically. We repeat for various $0
\le \varepsilon < 0.5$ and plot the resulting $D_e$ in
\figref{fig:23d_diag}(a) as a function of the solid volume
fraction $\phi ({\bf x}) =  \pi \varepsilon^d ({\bf x})$. The three-dimensional counterpart is shown in \figref{fig:23d_diag}(b).
\def \scl {.8}
\begin{figure}[ht]
\begin{center}
\psfragscanon
\psfrag{D3}[][][\scl][-90]{$D_{e}$} \psfrag{p3}[][][\scl]{$\phi$} 
\psfrag{D2}[][][\scl][-90]{$D_{e}$} \psfrag{p2}[][][\scl]{$\phi$} 
\psfrag{a}[][][\scl]{(a)} \psfrag{b}[][][\scl]{(b)} 
\includegraphics[width=.45\textwidth]{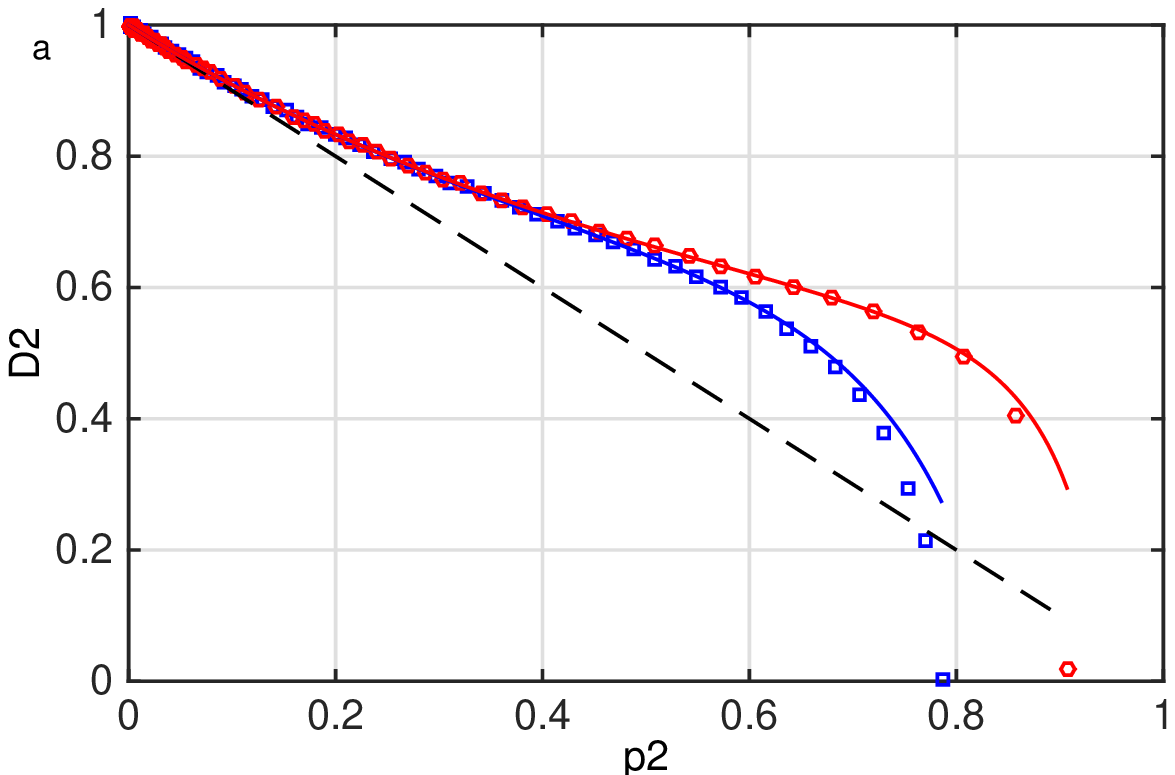}\quad \includegraphics[width=.45\textwidth]{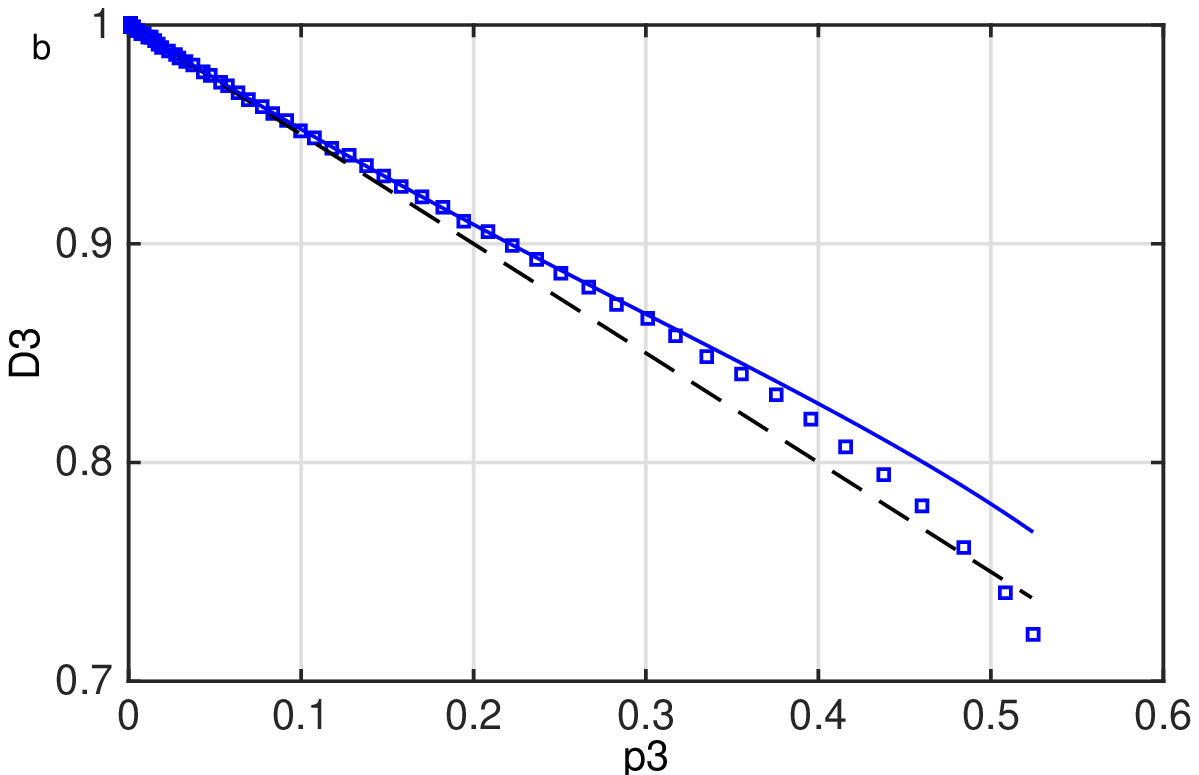}
\caption{Effective diffusion versus volume fraction $\phi$. Shown
  are the simulation results of $D_e$ in \eqref{Dx} (data points), Rayleigh's
  multipole method values (solid lines) and asymptotic result $\hat D_e$
  \eqref{point_Dcollective} (dash lines). (a) $d=2$: Square lattice
  [square data points and \eqref{multipoleS}] and hexagonal lattice
  (circle data points and \eqref{multipoleH}); asymptotic value $\hat D_e
  \sim 1- \phi$. (b) $d=3$: Cubic lattice (square data points and
  \eqref{multipoleC})  and asymptotic value $\hat D_e \sim 1-\phi/2$.} 
\label{fig:23d_diag}
\end{center}
\end{figure}

\begin{figure}[htb]
\begin{center}
\resizebox{0.5\textwidth}{!}{\input{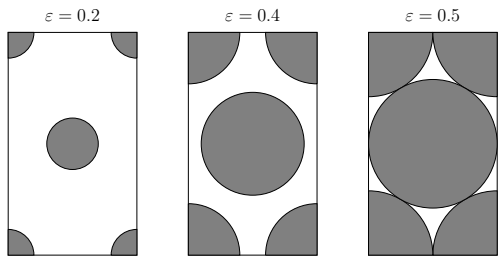}}
\caption{Periodic cell $\omega_v({\bf x})$ of the hexagonal
  configuration in two dimensions for different values of the
  obstacles' radius $\varepsilon$. The third cell corresponds to close
  hexagonal packing.} 
\label{fig:cellH}
\end{center}
\end{figure}

We next show with an example how, still using spherical obstacles, the
microscopic arrangement of the obstacles can alter the effective
diffusion coefficient, that is, two structures with the same porosity 
may have different $D_e$. 
Specifically, we consider the hexagonal packing configuration
in two dimensions. 
The periodic cell for such a configuration is $\omega = [-1/2,
  1/2]\times [-\sqrt 3/2, \sqrt 3/2]$ with one disk at the centre and
a quarter of a disk at each corner (see \figref{fig:cellH}).   
We solve this new cell problem and evaluate the
effective diffusion as a function of $\phi$ using \eqref{Dx}. The
resulting effective diffusion tensor is again a multiple of the
identity; we plot the scalar $D_e$ as a function of solid
volume fraction $\phi$ 
in \figref{fig:23d_diag}(a). We find that, for the same value of
$\phi$, there are differences in the effective diffusion coefficient in a 
square or hexagonal configuration, which become important for $\phi$
above 50$\%$.  

As mentioned in the introduction, the exact values of $D_e(\phi)$ in
ordered lattices can be computed 
using Rayleigh's multipole method, in the form of an infinite system
of algebraic equations involving image multipoles and lattice sums
\cite{Rayleigh:1892gq}. By truncating the infinite system, Rayleigh
obtained a closed-form approximation for a square lattice (S) and cubic lattice
(C). The corresponding result for a two-dimensional hexagonal lattice
(H) was derived in \cite{Johannesson:1996vt} as: 
\begin{subequations}
\label{multipole}
\begin{alignat}{2}
\label{multipoleS}
D_e^S (\phi) &= \frac{1}{1-\phi} \left ( 1 - \frac{2\phi}{1+\phi -
  0.3058 \phi^4} \right), &\qquad &(\phi < 0.7) \\ 
\label{multipoleH}
D_e^H (\phi) &= \frac{1}{1-\phi} \left ( 1 - \frac{2\phi}{1+\phi -
  0.07542 \phi^6} \right),  & &(\phi < 0.8)\\ 
\label{multipoleC}
D_e^C (\phi) &= \frac{1}{1-\phi} \left ( 1 - \frac{3\phi}{2+\phi -
  0.3914 \phi^{10/3}} \right), & &(\phi < 0.25) 
\end{alignat}
\end{subequations}
The values in parenthesis indicate the maximum volume fraction for
which the closed forms are valid
\cite{Johannesson:1996vt}.\footnote{The exact result of the multipole
  method for various geometries in two and three dimensions is
  computed in \cite{Johannesson:1996vt} by direct inversion of the
  matrix, for a growing number of multipoles (with the matrix becoming
  larger and larger) until the result converges.} 
These formulae are also plotted in  \figref{fig:23d_diag}, and
show good agreement with the multiple-scales result.

\subsection{Extension to non-periodic structures}
\label{sec:conformal}

Until now we have considered porous media with a simple regular
structure,
with the centers of the obstacles lying on a regular lattice, 
and accounted for macroscopic
gradients in porosity by allowing the radius of the obstacles to vary
slowly with position. 
In this section we consider an alternative way in which the porosity may
vary: we suppose that the obstacles are all the same size, but that the
number density of 
obstacles varies with position. This means that the obstacles can
no longer be arranged on a regular periodic lattice. However, we suppose that
their arrangement is locally a periodic lattice, but that the scale and
orientation of the lattice vary slowly with position.
To make this precise, we assume that the centers can be mapped  to
a regular  periodic   
lattice with a map that depends only on the slow scale. Effectively
this map defines (slow) curvilinear coordinates in which the structure is
periodic.
We use a recently developed generalization
of the multiple-scales method that can
handle such a microstructure  \cite{Richardson:2011cb}.

Thus we suppose there is exists a transformation ${\bf W}: \Omega \to
\Omega'$ 
mapping the centers of the obstacles ${\bf r}_i$ into a regular
lattice with periodicity $\delta$, as illustrated in
\figref{fig:periodicdomain}. We could now apply the method of multiple
scales in the transformed domain, averaging over the fast scale,
before inverting the transformation to write the homogenized equation
back in the original variables. However, since the fast and slow
scales are treated as independent, there is no point in transforming
the slow scale, only to invert the transformation later.

Instead we suppose that the solution $C$ is a function of the slow
scale ${\bf x}$ and the transformed fast scale 
\[
{\bf y}   =\frac{ {\bf W}({\bf x})}{\delta},
\]
and that these variables are independent. Then, since the microstructure is
periodic in ${\bf y}$, we assume that $C$ is periodic in ${\bf
  y}$, with period one.
We denote the unit cell in ${\bf y}$ by $\omega$ and the
obstacle by ${\bf y}  \in \omega_s({\bf r}_i)$, where
\begin{align*}
\omega_s({\bf r}_i)= \{ {\bf y}
\in \omega: \delta  {\bf y}  \in {\bf W}( B_\epsilon({\bf
  r}_i)) - {\bf W}({\bf r}_i) \}. 
\end{align*} 
The available volume is then
$\omega_v({\bf x})=\omega \setminus \omega_s ({\bf x})$. 

Using the chain rule spatial derivatives now transform according to
\begin{equation}
\frac{\partial }{\partial x_i} \quad \to \quad  \frac{\partial
}{\partial x_i} + \frac{1}{\delta} F_{ij} \frac{\partial }{\partial
  y_j},  
\end{equation}
where $F_{ij} = \partial W_j/\partial x_i$ or $F = J_{\bf W}^T$,
and we are using the summation convention for repeated indices.

The derivation of the homogenized equation is similar to the perfectly
periodic case. The result is again an advection--diffusion equation
for the concentration $c$ as in \eqref{sol_inc}, 
\begin{subequations}
\label{sol_poreavmap}
\begin{align} 
\label{solo4m}
\frac{\partial c}{ \partial t}   =   \nabla_{\bf x} \cdot
\left[ D_e({\bf x})  \nabla_{\bf x} c -    \frac{D_e({\bf x})
    \nabla_{\bf x} \psi ({\bf x}) } { \psi ({\bf x}) } c  \right], 
\end{align}
with a modified diffusion tensor
\begin{align}
\label{Dxm}
D_e({\bf x}) &= I_d - \frac{1 }{\psi({\bf x})} F({\bf x}) \left(
\int_{\omega_v ({\bf x}) }    J^T_{\bf \Gamma} ({\bf x}, {\bf y}) \, \ud
    {\bf y} \right) F^{-1} ({\bf x}). 
\end{align}
The function ${\bf \Gamma} = \Gamma_i$ now satisfies the following cell problem
\end{subequations}
\begin{subequations}
\label{cellproblem_def}
\begin{alignat}{2}
 \nabla_{{\bf y}} \cdot \left( F^T F  \, \nabla_{{\bf y}} \Gamma_i
 \right) &=  0 &\qquad    &{\bf y} \in \omega_v ({\bf x}), \\ 
F^T F  \, \nabla_{{\bf y}} \Gamma_i \cdot {\bf n}_{\bf y} & =F^T F\, {\bf
  n}_{\bf y} \cdot  {\bf e}_i &  &{\bf y} \in \partial \omega_v({\bf
  x}), \\ 
\Gamma_i & \quad  \textrm{periodic} & &\textrm{in} \ {\bf y}.
\end{alignat}
\end{subequations}
In general the cell problem \eqref{cellproblem_def} 
now depends on the slow scale ${\bf x}$ not only through
$\omega_v({\bf x})$ but also 
because $F$ depends on ${\bf x}$.

\subsubsection{Conformal maps}
Since the Jacobian matrix $F$ is dependent only on the slow scale ${\bf x}$ it
represents a constant linear transformation on the fast scale. Such a
transformation can always be written
\[ F = UDV\]
where $U$ and $V$ are real orthonormal  matrices and $D$ is a  diagonal matrix. 
In fact, in our application we expect $U$ and $V$ to have positive
determinant and therefore be rotations (so that we maintain the
right-handedness of the coordinate system).
Thus $F$ can be thought of as successive application of a rotation, a
stretch along 
the coordinate axes, and then another rotation.

In the particular case in which the stretch is isotropic, so that the
entries along the diagonal in $D$ are all equal, then $F$ reduces to
an isotropic stretch and a single rotation. In this case we  may write
$F = a R$, where $a\not = 0$ is a scalar stretch, and $R$ is a
rotation matrix. 
Maps ${\bf W}({\bf x})$ whose Jacobian $F$ have this property preserve
angles, that is they are conformal. 
Because the stretch is isotropic, spherical obstacles
remain spherical under such a transformation,
with the new radius of the original sphere
centered at ${\bf x}$ being 
\begin{equation}
\label{map_radi}
\varepsilon({\bf x}) = \frac{1}{\delta} \left \| {\bf W}({\bf x} + \epsilon)
- {\bf W}({\bf x})\right \|. 
\end{equation}

For conformal maps the cell problem simplifies considerably, both
because the sphere is mapped to a sphere, but also because
\[
F^T F = a^2 R R^T = a^2 I_d.
\]
Thus the cell problem 
\eqref{cellproblem_def} reduces to our original cell problem
\eqref{cellproblem}, with the slow scale felt only through the change
in obstacle radius with position via \eqref{map_radi}.

\begin{figure}[htb]
\begin{center}
\begin{overpic}[width=.4\linewidth,tics=10]{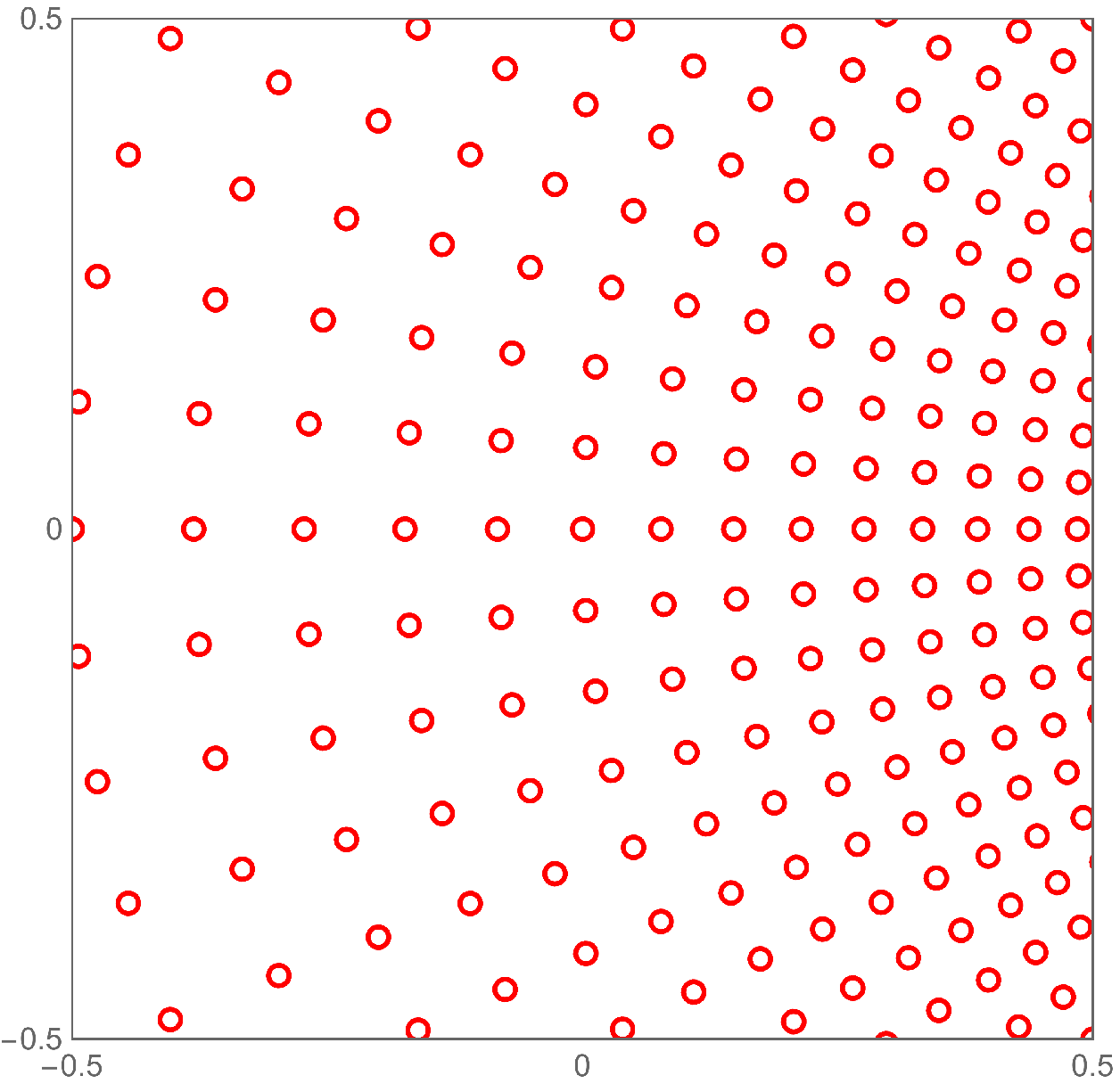} 
\put(-13,90){\small (a)}
\put(-1,50){\small $y$}
\put(50,-5){\small $x$}
\end{overpic}
\hspace{1cm}
\begin{overpic}[width=.4\linewidth,tics=10]{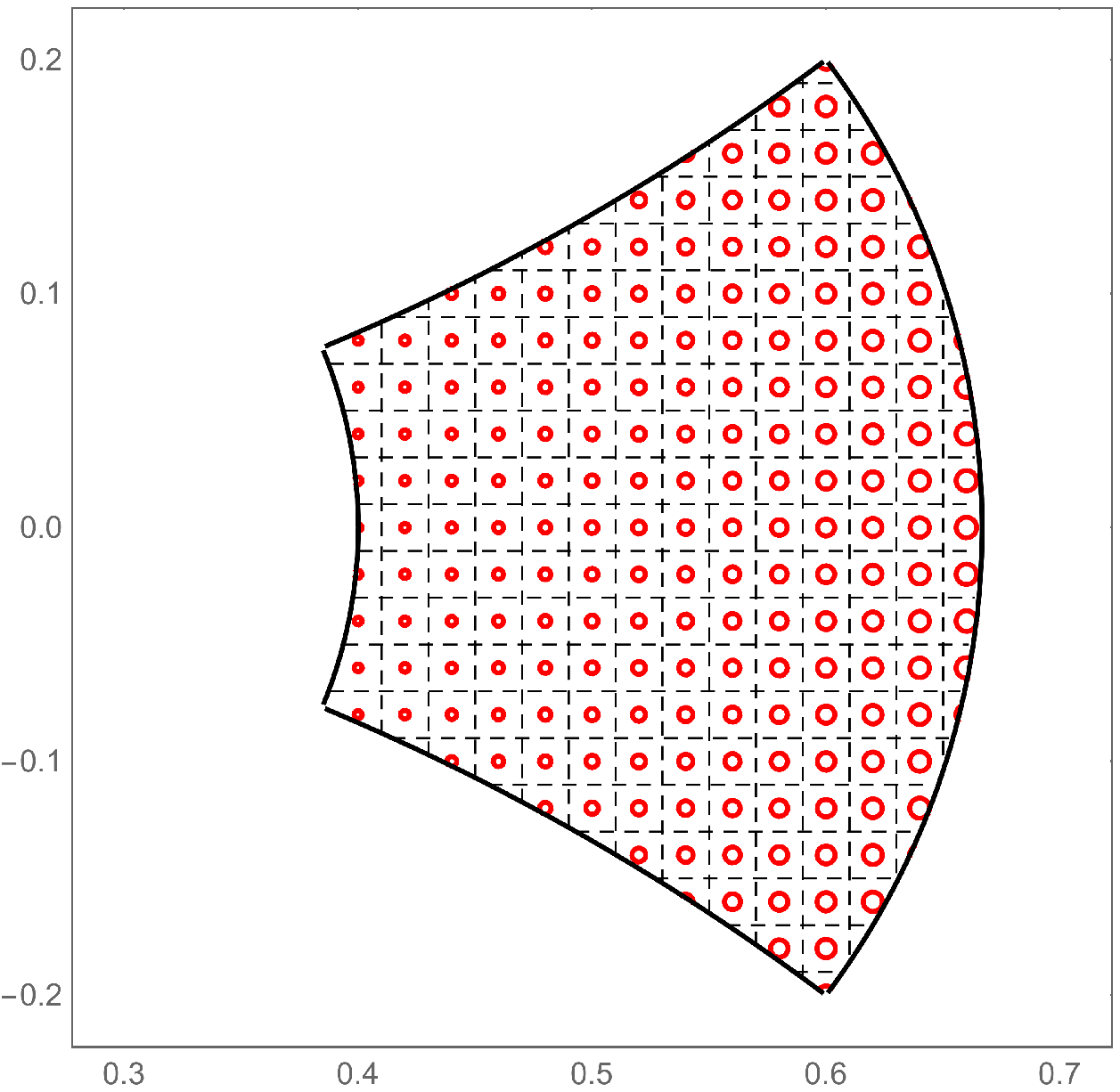} 
\put(-13,90){\small (b)}
\put(-3,50){\small $y'$}
\put(50,-5){\small $x'$}
\end{overpic}
\caption{Original domain $\Omega$ and the mapped domain $\Omega'$
  under the transformation \eqref{example_conf}. Parameters used are:
  $\delta = 0.02$, $\epsilon = 0.01$.} 
\label{fig:trans}
\end{center}
\end{figure}

In two dimensions, we may generate conformal maps by taking advantage
of complex variables. If we write $z = x + i y$ then any holomorphic 
function ${W: \Omega \subset \mathbb C \to \mathbb C}$ such that
$W'(z) \ne 0 $ for $z \in \Omega$ is a conformal map. 
An example that we will use in our simulations later is the  function
\begin{equation}
\label{example_conf}
z' = W(z) = \frac{1}{2-z},
\end{equation}
which is one-to-one and holomorphic  except at $z=2$.  In
particular, this defines a conformal map between $\Omega =
[-1/2,1/2]^2$ and $W(\Omega)$, with inverse $W^{-1} (z) = 2-1/z$.
In coordinate form,
\[
x' = \frac{2-x}{(2-x)^2 + y^2} \qquad y' =   \frac{y}{(2-x)^2 + y^2} .
\]
In \figref{fig:trans} we plot the original domain $\Omega$ and the
domain $\Omega' = W(\Omega)$. We see that in the transformed variables
${\bf x}'$ the centers of the obstacles lie on a regular square
lattice, but the radii of the obstacles vary with position, that is,
the structure is very similar to the one depicted 
in \figref{fig:periodicdomain}.  

\section{A random disordered porous medium}
\label{sec:stoch}

In this section we take a very different approach to model the same physical
problem. We assume that the solid matrix is still formed of
$N_s$ solid immobile non-overlapping spheres or disks of radius
$\epsilon$ at positions 
${\bf R}_i \in \Omega$, but now the ${\bf R}_i$ are randomly
distributed. The non-overlapping constraint means that the ${\bf
  R}_{i}$ cannot be distributed independently, but we suppose that
they are distributed identically, that is, the probability
distribution  function  is
invariant with respect to permutations of the particle labels
$1,\ldots,N_s$.

We suppose that there are  $N_m$ mobile solute particles at positions
${\bf X}_i (t)$, $i=1, \dots N_m$, at time $t$, which undergo Brownian
motion with constant  diffusion
coefficient $D_0$, which we may set to unity as before through
appropriate nondimensionalisation. In contrast to the previous
section, we relax the 
assumption that the solute particles are point particles and allow for
them to have a finite size. To keep things simple, we assume the solute
particles are also spherical and of radius $\epsilon_m \ll 1$ and
interact via hard-core elastic collisions with each other as well as
with the obstacles.  
We will use the same approach as in our previous works
\cite{Bruna:2012wu,Bruna:2012cg}, for which we require that the total
particle volume fraction is small so that three-particle
interactions and higher form a negligible fraction of state space and can be ignored. In terms of the parameters
introduced above, this means that $N_s \epsilon^d +  N_m \epsilon_m
^d\ll |\Omega| = 1$. 

Each solute particle evolves according to the overdamped Langevin stochastic differential equation 
\begin{align}
\label{sde_b}
{\rm d} {\bf{X}}_i  =  \sqrt{2}\, {\rm d}{\bf W}_i,
\end{align} 
where the ${\bf W}_i$ are $N_m$ independent $d$-dimensional standard
Brownian motions.  We suppose that the initial positions ${\bf  X}_i(0)$ are
random and identically distributed (that is, the probability
distribution is invariant with respect to a permutation of particle
labels). 

\subsection{The Fokker--Planck equation}
\label{sec:FP}

Let $P(\vec x, \vec r, t)$ be the joint probability density of the
$N_m+ N_s = N$ particles, where $\vec x = ({\bf x}_1, \dots, {\bf
  x}_{N_m})$ and $\vec r = ({\bf R}_1, \dots, {\bf R}_{N_s})$. We introduce the
$dN-$dimensional position vector $\vec \zeta = (\vec x,
\vec r)$ with the coordinates of the mobile
particles in the first $dN_m$ positions, and the obstacle 
positions in the remaining $dN_s$ positions. The probability density  
evolves according to the Fokker--Planck equation
\begin{subequations}
\label{twofp}
\begin{alignat}{2}
\label{twofp_eq}
\frac{\partial P}{\partial t} &=    \nabla_{\vec
  x}^2  P ,  \qquad  &(\vec x,\vec r)
&\in \Omega_\epsilon^{N}.
\end{alignat}
Although this looks like the standard diffusion equation
\eqref{origin_eq} we must remember that 
 \eqref{origin_eq} is solved  in the  low-dimensional physical space
 $\Omega_v \subset 
\Omega$ while  equation \eqref{twofp_eq} is solved in the high-dimensional
configuration space $\Omega_\epsilon ^N = \Omega^N \setminus \mathcal
B_\epsilon$, corresponding to $N$ copies of the physical domain
$\Omega$ minus the set of illegal configurations corresponding
to particles overlapping, 
\begin{equation*}
\mathcal B_\epsilon=\left \{\vec \zeta \in \Omega^{N}: \exists i\ne j
\quad \textrm{ 
s.t. } \quad  \| {\bo \zeta}_i -  {\bo \zeta}_j \| \le \epsilon_i +
\epsilon_j \right \}, 
\end{equation*}
where $\epsilon_i = \epsilon_m$ for $i\le N_m$ and $\epsilon_i =
\epsilon$ otherwise. It is convenient to introduce
$\epsilon_{sm}$  as the distance between centers at 
contact between an obstacle particle and a mobile particle,
$\epsilon_{sm} = \epsilon_m+ \epsilon$.  

On the collision surfaces
$\partial \Omega_\epsilon ^N$ we have the reflecting boundary
condition 
\begin{alignat}{2}
\label{twofp_bc} 
0 &= {\vec n}  \cdot \nabla_{\vec x} P , \qquad &
(\vec x, \vec r) & \in \partial  \Omega_\epsilon^N, 
\end{alignat} 
where $\vec n$ is the projection of the unit normal 
on the first $dN_m$ coordinates.  
\end{subequations}

Since $P(\vec x, \vec r, t) $ is invariant with respect to
permutations of the labels in $\vec r$,  the
marginal density functions of $P$ corresponding to fixing the position
of one obstacle and integrating over the positions of the remaining
obstacles are all identical, and given by
\begin{equation}
\label{s_def}
s({\bf x}) = \int_{\Omega_\epsilon^N} P(\vec x, \vec r, 0) \delta
({\bf r}_1 - {\bf x}) \, \ud \vec x \ud \vec r.
\end{equation}
This gives the 
probability of finding an obstacle in a given position, that is, 
it is the obstacle population density. 
 Since $|B_\epsilon|$ is the volume of one
obstruction, the local volume concentration of obstacles is
\begin{equation}
\label{obstacle_density}
\phi({\bf x}) = N_s |B_\epsilon| s({\bf x}) = \Phi \, s({\bf x}),
\end{equation} 
where, as before, $\Phi$ is the  average solid volume fraction.

In order to compare the Fokker--Planck equation \eqref{twofp} to the model we
have derived via multiple scales, we need an equation for the marginal
solute density $p({\bf x},t)$ defined on the physical domain
$\Omega$. As with  the obstacles, since the $N_m$ mobile particles are
identically distributed, $P$ is invariant to permutations of their
labels and we can define the solute  population density as 
\begin{equation}
\label{p_def}
p({\bf x},t) = \int_{\Omega_\epsilon^N} P(\vec x, \vec r, t) \delta
({\bf x}_1 - {\bf x}) \, \ud \vec x \ud \vec r. 
\end{equation}
Then the concentration of mobile particles is $c_m ({\bf x}, t) = N_m p({\bf x},t)$ and the normalized concentration (used in \S\ref{sec:determ}) is $c = c_m / \int_\Omega c_m \ud {\bf x} \equiv p$.

The procedure we adopt to derive an equation for $p$ from
\eqref{twofp} 
is a systematic asymptotic expansion as $N_s \epsilon^d + N_m \epsilon_m^d \rightarrow0$. We use matched asymptotic expansions in configuration space, with
an outer region in which particles are well-separated, and an inner
region in which two particles are close together. In contrast with
other approaches this systematic procedure does not require a closure
assumption. 

The derivation is
analogous to that presented in our previous work
\cite{Bruna:2012wu}, and we omit the details. In fact, the problem
studied here can be regarded as a 
particular case of the model for two species of diffusing and
interacting particles presented in \cite{Bruna:2012wu} if we formally set the
diffusivity 
of one of the species to zero. Although the derivation should really
be modified in this limit, the result is identical to equation (22a) in
\cite{Bruna:2012wu}, which, in our notation, reads 
\begin{subequations}
\label{asym_density}
\begin{align}
\label{finalmodelc_b_obs0}
\frac{\partial p }{\partial t}  = \nabla_{\bf x}  \cdot
\big [ \hat D_e ({\bf x})  \nabla_{\bf x}  p   -  \hat {\bf v} ({\bf
    x})  p  \big], 
\end{align}
where the diffusion and drift coefficients are
\begin{align}
\label{Dcollectiveb}
\hat D_e ({\bf x}) &= 1 + (N_m-1) \frac{(d-1) \pi}{ 2^{d-1} d}
\epsilon_m^d  p({\bf x}) - N_s  \frac{2 \pi}{d} \epsilon_{sm}^d s({\bf
  x}),\\ 
\label{Vcollectiveb}
\hat {\bf v}({\bf x}) &= - N_s \frac{2 (d-1) \pi}{d} \epsilon_{sm}^d \nabla_{\bf x}  s({\bf x}),
\end{align}
\end{subequations}
for $d=2,3$. The effective diffusion coefficient of solute particles
has three components: the base molecular diffusion, an enhanced diffusion due to
collective motion of finite size solute particles, and reduced diffusion due to
excluded-volume interactions with the obstacle particles. The
effective advection $v({\bf x})$ is due to the gradient in the
distribution of obstacles; it indicates that the particles are advected
towards regions with fewer obstacles. Note that there is no
system-size expansion in (\ref{asym_density}): the equations are
equally valid with small numbers of solute particles or obstacles (for
example, the collective term disappears if there is only one solute
particle, as it should).

\subsection{Model for point particles diffusing  in a stochastic porous medium}
\label{sec:stoch_limit}

The two approaches we have taken each have their strengths and
weaknesses. The multiple-scales approach can handle arbitrary volume
fraction of obstacles, but is limited to locally periodic structures
and the diffusion of point
particles of solute. The Fokker--Planck approach can handle arbitrary
obstacle configurations and finite-size solute particles, but can only
be reduced to a low-dimensional effective diffusion equation in the
limit of small volume fraction.

In order to compare the two approaches later, we set $\epsilon_m = 0$ 
in \eqref{asym_density} to consider point solute particles.
 The effective diffusion and drift coefficients
reduce to  
\begin{subequations}
\label{point_eff}
\begin{align}
\label{point_Dcollective}
\hat D_e ({\bf x}) &= 1 - \frac{\Phi}{(d-1)}  s({\bf x}) = 1- \frac{\phi({\bf x})}{(d-1)},\\
\label{point_Vcollective}
\hat {\bf v} ({\bf x}) &= - \Phi \nabla_{\bf x}  s({\bf x}) = -\nabla_{\bf x}  \phi({\bf x}),
\end{align}
where $\Phi = \frac{2 (d-1) \pi}{d} \epsilon^d N_s$. We plot $\hat
D_e$ given by \eqref{point_Dcollective} in \figref{fig:23d_diag} with dashed
lines. We observe that it agrees with the multiple-scales value $D_e$
for small volume fractions $\phi$ (we will show this formally in \S
\ref{sec:deter_limit}). We note that, in the cubic obstacle 
configuration ($d=3$), the asymptotic value  $\hat
D_e$  is a good approximation to
$D_e(\phi)$ throughout the whole  range of valid porosities.  

For point solute particles, the stationary density $p_\infty({\bf x})$
takes a very simple form. Substituting \eqref{point_eff} in
\eqref{finalmodelc_b_obs0} and imposing no-flux boundary conditions,
we find 
\end{subequations}
\begin{equation}
\label{stationary_points}
p_\infty({\bf x}) = \kappa \left[ 1- \frac{\phi({\bf x})}{(d-1)}
  \right]^{d-1}
\sim  \ 1- \phi({\bf x}) = \psi({\bf x}),
\end{equation}
where $\kappa$ is the normalization constant. This corresponds to the
uniform measure in the available space,
and is consistent with the
stationary density found via multiple scales.

\subsection{Diffusion through moving obstacles is easier than through
  fixed obstacles} 
\label{sec:moving}

Before moving on to compare our different approaches, we briefly examine  how the effective transport properties of the solute particles change when
the obstacles themselves are 
allowed to diffuse, with molecular diffusion $D_s$ (in our dimensionless setting this represents the ratio of the diffusion
coefficient of the obstacles to that of the solute). This would be relevant in biological applications such as the diffusion through biological tissues or the cytoplasm, where one is interested in the diffusion of small molecules through an environment containing large macromolecules \cite{Novak:2009ck}. These macromolecules constitute the ``solid phase'', which is sometimes considered as a porous structure since its dynamics are much slower than the smaller diffusing particles. Using the general model in \cite{Bruna:2012wu} the coefficients
in \eqref{point_eff} change to 
\begin{align}
\label{move_eff}
\hat D_e ({\bf x},t)  = 1- \frac{1}{1+D_s}\frac{\phi({\bf x},t)}{(d-1)},\qquad 
\hat {\bf v}({\bf x},t) = - \frac{(d-1) + d D_s}{(d-1) (1+D_s)} \nabla_{\bf x}  \phi({\bf x},t),
\end{align}
 for $d=2,3$. Setting $D_s=0$ in \eqref{move_eff}  recovers
the expressions in \eqref{point_eff} as expected. From
\eqref{move_eff} we see that the faster the obstacles move (the larger
$D_s$), the less they impede the diffusion of the point solute
particles, with their effect disappearing completely as $D_s\to
\infty$. On the other hand, the 
larger $D_s$ is, the larger the coefficient in front of the drift
term $\hat {\bf v}({\bf x})$. This drift does not disappear in the
limit, with $\hat {\bf v} = -d \nabla_{\bf 
  x}  \phi / (d-1)$ as $D_s\rightarrow \infty$. 
Of course, since the obstacles are much larger than the solute
particles, we would expect that they diffuse more slowly, i.e. that
$D_s<1$ in any practical situation.

\section{Comparison between methods}
\label{sec:comparision}

In this section we compare the macroscopic models for diffusion in a
porous medium of variable porosity which we 
derived via multiple scales in \S\ref{sec:determ}
and using the Fokker--Planck approach in \S\ref{sec:stoch}. 
As we mentioned above, 
the Fokker--Planck approach can only
be systematically 
reduced to a low-dimensional effective diffusion equation in the
limit of small volume fraction. We observed in \figref{fig:23d_diag}
 that the multiple-scales-derived diffusion coefficient seems
 numerically to asymptote to the Fokker--Planck-derived diffusion
 coefficient in this limit. In the next section we show that this is
 indeed the case, by considering the asymptotic
solution to the multiple-scales model in the limit of low obstacle
volume fraction $\Phi$.  
We then compare our effective equations with each other and with
direct numerical simulations in a variety of test problems.

\subsection{Model for an ordered porous medium with low porosity}
\label{sec:deter_limit}

We consider the model \eqref{sol_poreav} in the limit of low volume
fraction $\Phi$. This means that 
 the local volume fraction
$\phi({\bf x})$ is also small (almost everywhere), corresponding 
to small relative obstacle radius $\varepsilon({\bf x}) \ll 1$. In
this limit, the cell problem \eqref{cellproblem} can be 
solved explicitly. 

Since ${\bf x}$, and hence $\varepsilon ({\bf x})$, are constants as far
as the cell problem \eqref{cellproblem} is concerned, we can look for
an asymptotic solution to \eqref{cellproblem} in terms of the small
parameter $\varepsilon ({\bf x})$. Consider, say, the first component
of the vector function ${\bo \Gamma}$, which satisfies 
\begin{subequations}
\label{cellproblem_1}
\begin{alignat}{2}
\nabla^2_{{\bf y}} \Gamma_1 &=  0 &\qquad   & {\bf y} \in \omega_v ({\bf x}), \\
\nabla_{{\bf y}} \Gamma_1 \cdot  {\bf y}  & = y_1 & & \| {\bf y}\| = \varepsilon({\bf x}), \\
\Gamma_1 & \quad  \textrm{periodic} & &\textrm{in} \ {\bf y}.
\end{alignat}
\end{subequations}
We use the method of matched asymptotic expansions, supposing that the
unit cell $\omega_v ({\bf x})$ can be divided into 
two regions: an inner region in which \mbox{$\|{\bf y}\| \sim  \mathcal O(
\varepsilon )$},  and an outer region in which $\|{\bf y}\| \gg
\varepsilon$.  In the inner region, we set ${\bf y} = \varepsilon({\bf
  x}) {\bf Y}$ and define $\gamma_1({\bf x}, {\bf Y}) = \Gamma_1({\bf
  x}, {\bf y})$ to give 
\begin{subequations}
\label{cellproblem_as2}
\begin{alignat}{2}
\nabla^2_{{\bf Y}} \gamma_1 &=  0 &\qquad   &  \\
\label{cellproblem_bc11}
\nabla_{{\bf Y}} \gamma_1 \cdot  {\bf Y}  & = \varepsilon Y_1 & & \textrm{on} \quad \| {\bf Y}\| = 1,
\end{alignat}
\end{subequations}
with a matching condition as $\|{\bf Y}\| \rightarrow \infty$. 
Expanding $\gamma_1({\bf x},{\bf
  Y}) = \gamma_1^{(0)}({\bf x}, {\bf Y}) + \varepsilon
\gamma_1^{(1)}({\bf x}, {\bf Y}) + \cdots$ gives that  the leading-order inner solution $\gamma_1^{(0)}$ is simply a constant in $\bf
Y$, whence the leading-order outer solution is also constant. 
At first order in $\varepsilon $ we find 
\begin{subequations}
\label{cellproblem_as3}
\begin{alignat}{2}
\nabla^2_{{\bf Y}} \gamma_1^{(1)} &=  0 &\qquad   &  \\
\nabla_{{\bf Y}} \gamma_1^{(1)} \cdot  {\bf Y}  & = Y_1 & &
\textrm{on} \quad \| {\bf Y}\| = 1. 
\end{alignat}
\end{subequations}
Using polar ($d=2$) or spherical ($d=3$) coordinates, we look for a
solution to \eqref{cellproblem_as3} of the form $\gamma_1^{(1)} =
f({\bf x},R) \cos \theta$ ($d=2$) or $\gamma^{(1)} = f({\bf x},R) \sin \theta
\cos\varphi$ ($d=3$), where $Y_1 = R \cos \theta$ when
$d=2$ and $ Y_1 = R \sin\theta \cos \varphi$ when $d=3$. We find that 
\begin{equation}
\label{pre_A}
f({\bf x}, R) =  A({\bf x}) R +  \frac{[A({\bf x}) - 1]}{(d-1)R^{d-1}},
\end{equation}
for an unknown function $A({\bf x})$. Since the leading-order outer
solution is constant in ${\bf y}$, matching gives $A({\bf x}) \equiv
0$. Thus
\begin{equation}
\gamma_1^{(1)}({\bf x}, {\bf Y}) = - \frac{1}{(d-1)} \frac{Y_1}{\|
  {\bf Y}\|^d} + \hat{\gamma}_1^{(1)}({\bf x}).\label{gamma1inner}
\end{equation}
Thus the first non-constant term in the outer expansion is ${\mathcal
  O}(\varepsilon^d)$. Matching with the outer solution gives that 
\begin{equation}
\label{sol_asy}
\Gamma_1({\bf x}, {\bf y}) \sim \mbox{constant}  - \frac{
  \varepsilon^d ({\bf x}) }{(d-1)} \frac{y_1}{\| {\bf y}\| ^d} + \cdots, 
\end{equation}
 as $\|{\bf y}\|\to 0$.

We can now use this asymptotic behavior to determine the outer
solution at this order. However, it is possible to determine the
effective diffusion coefficient, which is our primary aim, with the
information we already have. Since the integrals we have to evaluate
in \eqref{Dx} are all derivatives with respect to some component of
${\bf y}$, by integrating with respect to this component first we turn
the volume integral over the unit cell $\omega_v({\bf x})$ into
surface integrals over the exterior periodic boundaries and the
interior boundary with the solid 
obstacle.  The contributions from the exterior boundaries cancel
due to periodicity, while on the interior boundary we can use the
asymptotic solution
\eqref{gamma1inner}. 
The result  is 
\[\int_{\omega_v({\bf x})} J_{\bo \Gamma}^T \, \ud
{\bf y} = \frac{2\pi \varepsilon^d({\bf x})}{d}\,  \delta_{ij}. 
\]
Thus as $\varepsilon \to 0$, 
$D_e$ is a 
scalar multiple of the identity, equal to 
\begin{subequations}
\begin{equation}
\label{difftensorhat}
D_e ({\bf x})  \sim 1 - \frac{1}{ \psi({\bf x})} \frac{2\pi}{d}  \varepsilon^d ({\bf x})  \sim 1 - \frac{2\pi}{d}  \varepsilon^d ({\bf x}) = 1 - \frac{\phi({\bf x}) }{(d-1)},
\end{equation}
since  $\psi ({\bf x}) = 1- \phi({\bf x}) = 1 - 2(d-1)\pi
\varepsilon^d /d \sim 1$ at leading order.  
As expected, this result agrees with the asymptotic value
\eqref{point_Dcollective} obtained with the Fokker--Planck approach.  

The drift term in the multiple-scales homogenized equation
\eqref{sol_inc} is ${\bf v} ({\bf x}) =  D_e({\bf x}) \nabla_{\bf x}
\psi ({\bf x}) / \psi ({\bf x})$. Using \eqref{difftensorhat}, we
obtain 
\begin{equation}
\label{drift_as}
{\bf v}({\bf x}) = - \frac{D_e({\bf x}) \nabla_{\bf x} \phi ({\bf x}) }{ 1- \phi ({\bf x})} \sim - \frac{d-1-\phi}{(d-1) (1-\phi)} \nabla_{\bf x} \phi ({\bf x}) \sim - \nabla_{\bf x} \phi ({\bf x}).
\end{equation}
This asymptotic value also agrees with the drift obtained in the reduced
Fokker--Planck model; see \eqref{point_Vcollective}.  
\end{subequations}

Finally we comment that the nature of the calculation above makes it
clear that the configuration of the inclusions does not affect the
diffusion coefficient at this order, since the dominant contribution
comes from the solution of the cell problem in the inner region, which
has no information about the position of the inclusion(s) in the unit cell.

\subsection{Numerical simulations}
\label{sec:simulations}

The aim of this section is to compare the two models for diffusion in
porous media against each other as well as against numerical
simulations for the full problem and stochastic
simulations of the particle system. 

First, we consider the mean squared displacement  of
particles diffusing in two homogeneous porous media with the same
porosity, namely a deterministic structure with a square lattice of obstacles and a random structure with obstacle configurations drawn from a uniform distribution with non-overlapping constraints. 

Second, we consider the spreading out of a localized initial
concentration of particles in porous media with
gradients in porosity. Again we consider locally periodic structures
accessible to the multiple-scales analysis and random structures with
the same (ensemble) average porosity. 

All the simulations are made in a two-dimensional unit square domain $\Omega$ with $N_m$ point mobile particles and $N_s$ hard-disk obstacles of constant radius $\epsilon$. When the porous structure is random with probability law $s({\bf x})$, a new configuration of obstacles is generated for every new run.

\subsubsection{Effective diffusion coefficient via the mean squared
  displacement} 
\label{sec:self-diffusion}

In this section we compare the diffusion coefficient computed from
simulations of the discrete stochastic system to the effective
diffusion coefficient obtained in the previous sections, either from
multiple scales $D_e$ \eqref{Dx} if the obstacles are placed in a
regular structure, or  from the Fokker--Planck description $\hat D_e$
\eqref{point_Dcollective} if the obstacles are randomly distributed
with density $s({\bf x})$. In particular, we consider two porous media with uniform 
porosity $\Phi$ (so that $s({\bf x}) \equiv 1$ so that the drift is ${\bf v} \equiv
0$) and the same number $N_s$ of obstacles: (i) a square lattice configuration, and (ii) a random uniform configuration of obstacles with non-overlapping constraints.

The numerical value of the diffusion coefficient is obtained from the mean-square displacement, using the relation $\langle r^2 (t) \rangle = 2d D_e t$ as $t \to \infty$. To evaluate the mean-square displacement, we run $M = 1000$ runs with $N_m = 100$ mobile particles, and compute $\langle r^2 (t) \rangle = \frac{1}{M N_m} \sum_{k=1}^{M} \sum_{i = 1}^{N_m} 
\| {\bf X}_i^{(k)}(t) - {\bf X}_i^{(k)}(0) \| ^2$. Here ${\bf X}_i^{(k)}(t)$ is
the position of the $i$th particle in the $k$ realization at time $t$.\footnote{Since every realization is done with $N_m$ point particles, this is equivalent to averaging over particle trajectories and regenerating the solid matrix every $N_m$ realizations).} 

Our stochastic simulations are performed integrating
Eq. \eqref{sde_b}  using an Euler--Maruyama scheme, with reflective boundary conditions between mobile particles and obstacles ($\partial \Omega_s$) and periodic boundary conditions on the outer boundary ($\partial \Omega$).  The reflective boundary conditions between the mobile point particles and the obstacles are implemented similar to as in \cite{Bruna:2012wu}, namely, the distance that a particle has travelled (illegally) inside an obstacle is reflected back into the domain $\Omega_v$. To do that, we compute the point on the obstacle boundary where the particle penetrated, and compute a particle--wall elastic collision on that point. The integration timestep $\Delta t$ must be chosen carefully so that virtually no collisions are missed. A convergence study is shown in the Appendix and based on this, we have used $\Delta t =  1.2434 \cdot 10^{-6}$ in the results presented below. 

We perform experiments with the periodic and random porous media at porosities $\Phi = 10, 20 $ and 30\%. From \figref{fig:23d_diag}(a), we expect differences between the random and periodic porous media to become apparent from $\Phi = 20\%$. The question is whether the discrepancy between $D_e$ and $\hat D_e$ is real (due to the structure) or artificial (due to the nature of the asymptotic approximation in obtaining $\hat D_e$). 

We plot the mean-square displacement $\langle r^2 (t) \rangle$ against time in  \figref{fig:msd_simul}(a) for the square lattice (solid lines with error bars) and the random structure (dashed lines with crosses and error bars). (The error bars indicate the 95\% confidence interval, or $\pm 1.96$SD values, of each data point,  and are barely discernible.) As expected, $\langle r^2 \rangle$ increases linearly with time, with a slope that decreases with $\Phi$. 

\def \scc {0.65}
\def \scl {.8}
\begin{figure}[htb]
\unitlength=.8cm
\centering
\psfragscanon
\psfrag{t}[][][\scl]{$t$} 
\psfrag{phi}[][][\scl]{$\Phi$} 
\psfrag{p1}[][][\scl]{$10\%$} 
\psfrag{p2}[][][\scl]{$20\%$} 
\psfrag{p3}[][][\scl]{$30\%$} 
\psfrag{phi1}[][][\scc]{$10\%$} 
\psfrag{phi2}[][][\scc]{$20\%$} 
\psfrag{phi3}[][][\scc]{$30\%$} 
\psfrag{datada1}[][][\scc]{$\hat D_e$ th.\ } 
\psfrag{datada2}[][][\scc]{$\hat D_e$ sim.} 
\psfrag{datada3}[][][\scc]{$D_e$ th.\ } 
\psfrag{datada4}[][][\scc]{$D_e$ sim.} 
\psfrag{msd}[][][\scl]{$\langle r^2 \rangle$} 
\psfrag{msd4t}[][][\scl]{$\langle r^2 \rangle/(4 t)$} 
\psfrag{a}[][][\scl]{\,(a)} 
\psfrag{b}[][][\scl]{\,(b)} 
\psfrag{datadata0}[][][\scc]{simulation} 
\psfrag{datadata1}[][][\scc]{$\hat D_e$} 
\psfrag{datadata2}[][][\scc]{$D_e$}
\psfrag{datadata3}[][][\scc]{$\Phi = 0$}  
\includegraphics[width=.47\linewidth]{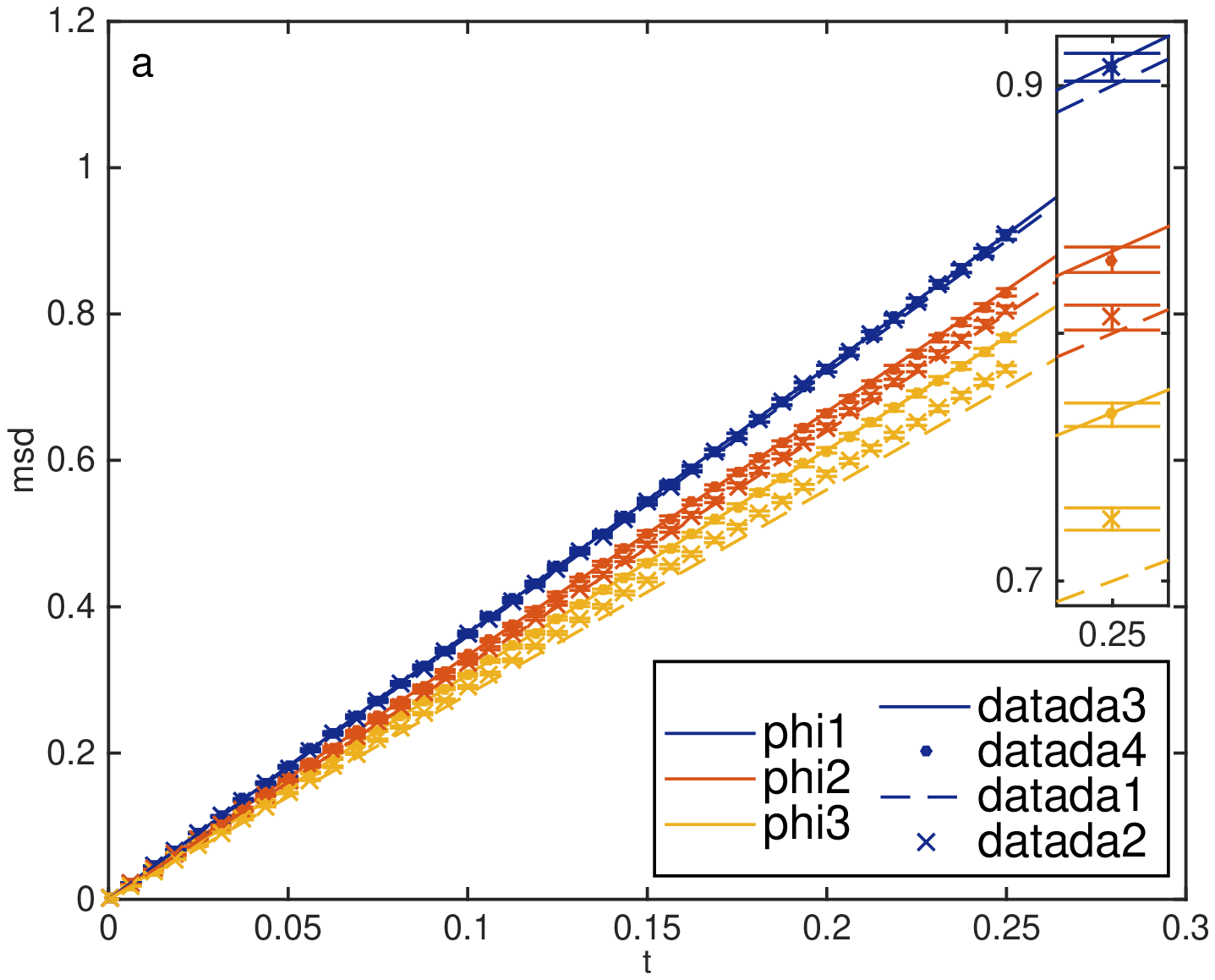}  \quad
\includegraphics[width=.47\linewidth]{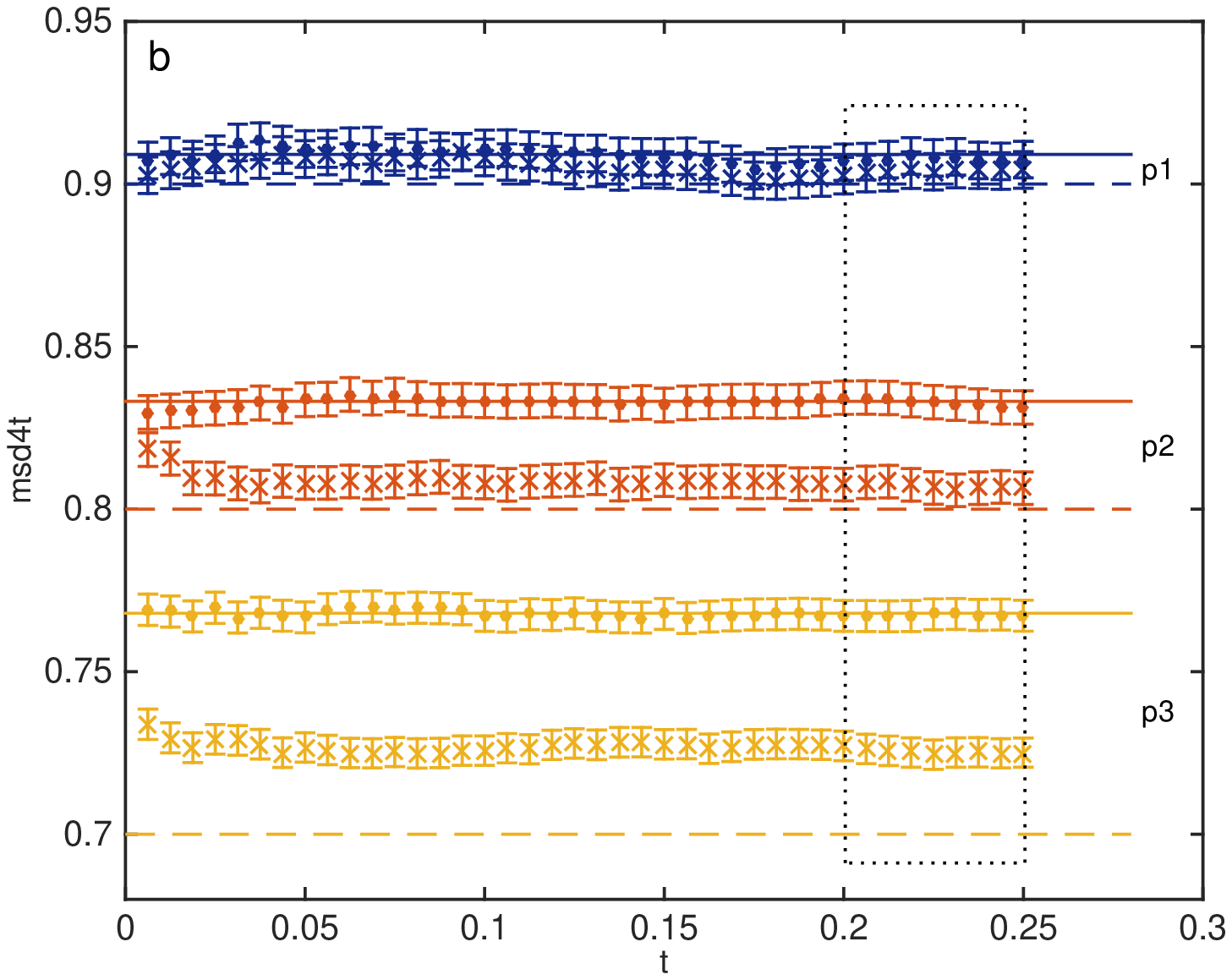} 
\caption{(a) Mean squared displacement $\langle r^2 \rangle$ as a function of time $t$ for diffusion of point particles in the presence of obstacles. The obstacles are arranged on a square lattice (error bars) or uniformly distributed (error bars with crosses) and at volume fraction $\Phi = 10, 20, 30\%$. The theoretical curves using $D_e$ (Eq.  \eqref{Dxm}) and $\hat D_e$ (Eq. \eqref{point_Dcollective}) are shown with solid lines and dashed lines, respectively. 
(b) The same data replotted as $\langle r^2 \rangle /(4t)$, to demonstrate that by the final simulation time the numerical diffusion coefficient has converged. The data points  inside the dashed black rectangle are used to compute the numerical value of diffusion. In both plots, the error bars indicate the 95\% confidence interval. 
Parameters used for the simulations are: $N_m = 100$, $M = 1000$ (a new obstacle configuration was generated at every new run in the random case), $\Delta t =  1.2434 \cdot 10^{-6}$, $\epsilon = 0.0126$, and $N_s = 200, 400, 600$ for $\Phi = 0.1, 0.2, 0.3$ respectively.} 
\label{fig:msd_simul}
\end{figure}

The diffusion coefficient is given by $\lim_{t \to \infty} \langle r^2 (t) \rangle/(4 t)$. To check that we have run the simulation for long enough and rule out any anomalous transient diffusion, in \figref{fig:msd_simul}(b) we plot $ \langle r^2 (t) \rangle/(4 t)$ \cite{Berezhkovskii:2014gb}. This type of plot highlights any time dependence in the diffusion coefficient \cite{Saxton:1994hk}. We observe that, for all curves in \figref{fig:msd_simul}(b), the curves have slope 0 after the first $t=0.05$ and hence have converged. To evaluate the diffusion coefficient, we average over the last $\Delta t = 0.05$ of each simulation (data points marked with a dashed rectangle in \figref{fig:msd_simul}(b)). From \figref{fig:msd_simul}(b) and the estimated values (data not shown), we note that: (i) For the square lattice case, theory (eq. \eqref{Dxm}, solid lines) and simulation results (solid error bars) for $D_e$ agree very well, as expected since $D_e$ from \eqref{Dxm} is exact. (ii) The random media simulation results (error bars with crosses) agree well with the asymptotic value $\hat D_e$ (eq. \eqref{point_Dcollective}, dashed lines) for $\Phi = 0.1, 0.2$ but there is a significant discrepancy for $\Phi = 0.3$. 
(iii) The difference between the regular and random porous media is within the error bars for $\Phi = 0.1$ but becomes apparent for $\Phi = 0.2, 0.3$. 

Interestingly, while the multiple-scales effective
diffusion coefficient $D_e$ does a better job for the period medium (as we might expect), the Fokker--Planck effective diffusion  coefficient  $\hat D_e$ seems to do a slightly better job for the random medium. This was not at all obvious, since this coefficient is only the leading term in an expansion as $\Phi \rightarrow 0$ (while the multiple-scales coefficient is valid for all $\Phi$). It seems that, for a given obstacle volume fraction $\Phi$, random porous media may have a slightly lower diffusion coefficient than periodic ones. Because $\hat D_e$ is an asymptotic expansion, we can use the order of the next term to estimate the error. The next term in the asymptotic expansion of $\hat D_e$ is $O\left((2\epsilon)^4 N_s^2, (2\epsilon)^3 N_s \right)$ \cite{Bruna:2012cg}, which gives 0.065 and 0.1459 for $\Phi = 0.2$ and 0.3 respectively. The discrepancies in \figref{fig:msd_simul}(b) between the asymptotic values $\hat D_e$ and the simulation results for $\Phi = 0.2$ and 0.3 are 0.0103 and 0.0257, respectively. In other words, the effective diffusion $\hat D_e$ does better than expected from the asymptotic error bounds. 

\subsubsection{Diffusion in a gradient of porosity}
\label{sec:gradient}

For our second model comparison, we consider a porous medium with a
non-uniform porosity. As before, we consider both a locally periodic
array of obstacles of constant radius $\epsilon$, and a random array
of obstacles giving the same local porosity.

For the locally periodic structure we use the arrangement
illustrated in \figref{fig:trans}(a).
 The obstacles have a fixed radius $\epsilon =0.01$, and 
the average volume fraction of obstacle is \mbox{$\Phi = N_s \pi \epsilon^2
= 0.059$}.  
To generate a random periodic medium with the same (ensemble) average
local porosity we need to determine  the probability density function
of obstacle position $s({\bf x})$, which we have seen is related to the
 volume concentration of obstacles by 
\begin{equation}
\label{obstacle_density2d}
\phi({\bf x}) =  \Phi s({\bf x}).
\end{equation}
This is easily found once we have determined the variable obstacle
 radius $\varepsilon({\bf x})$ in the mapped multiple-scales
domain (shown in  \figref{fig:trans}(b)).

\begin{figure}[htb]
\begin{minipage}[c]{0.46\linewidth}
\centering
\begin{overpic}[width=.8\linewidth,tics=10]{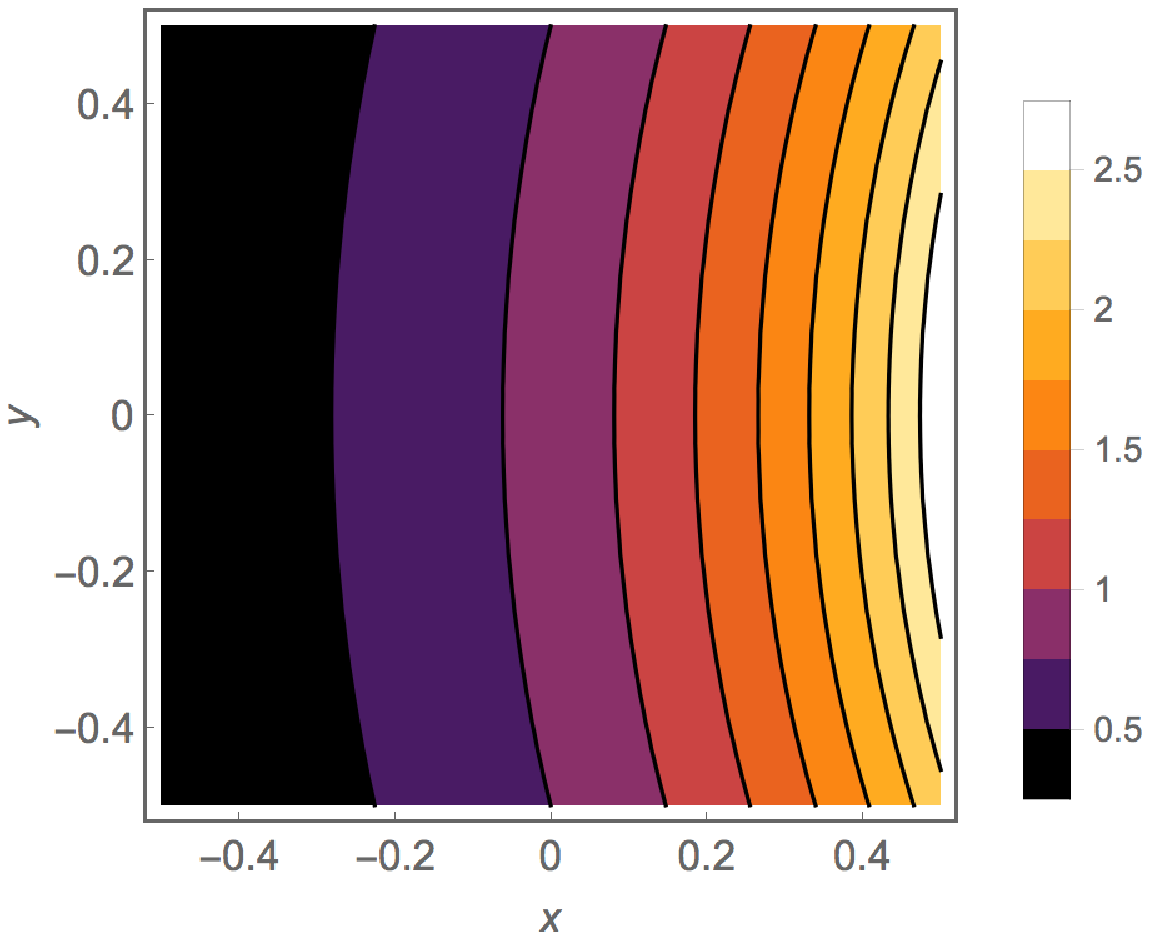} 
\put(-5,75){\small (a)}
\end{overpic}
\end{minipage}
\
\begin{minipage}[c]{0.46\linewidth}
\centering
\begin{overpic}[width=.7\linewidth,tics=10]{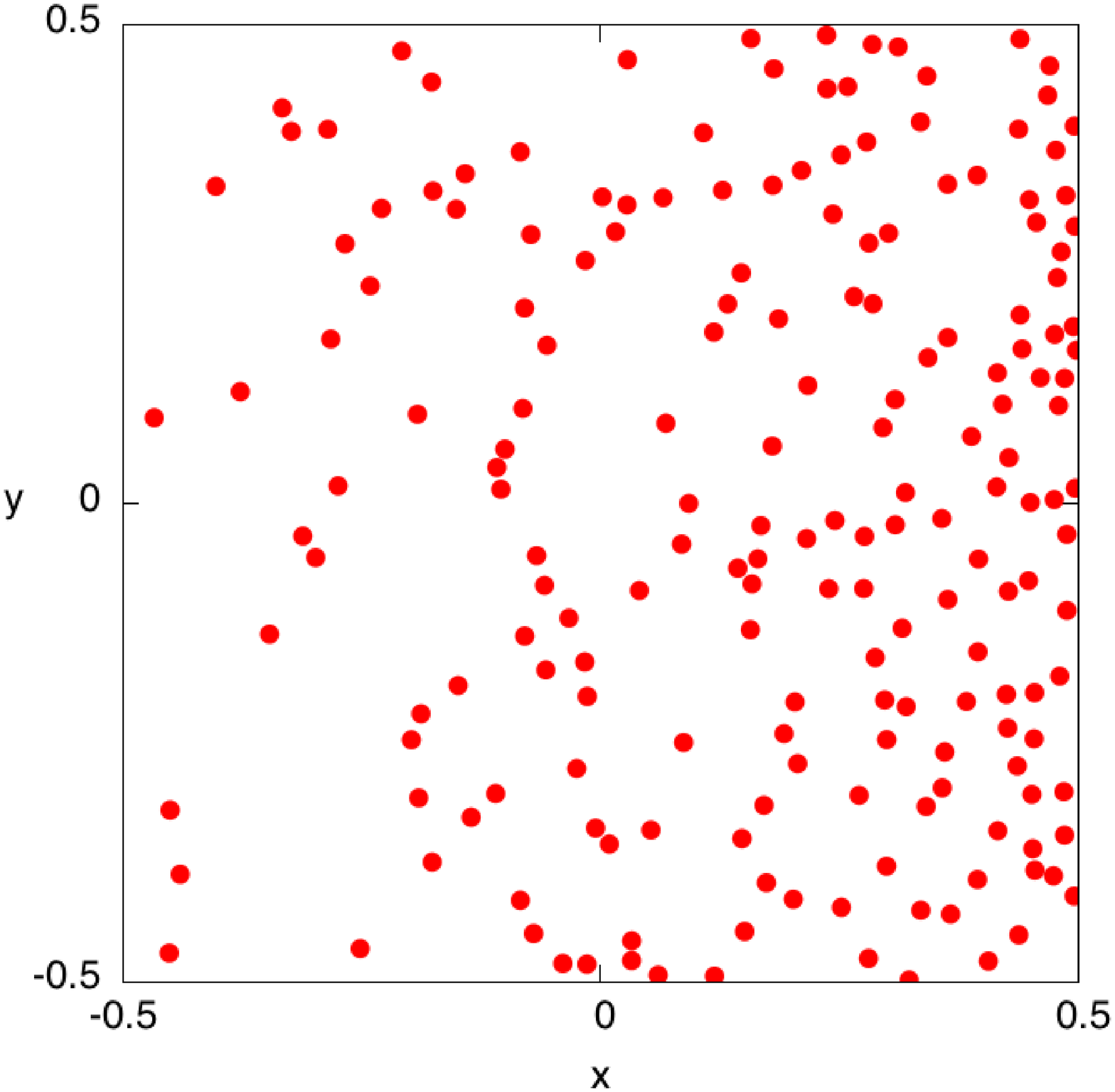} 
\put(-10,90){\small (b)}
\end{overpic}
\end{minipage}
\caption{(a) Density of obstacles $s({\bf x})$ corresponding
  to the  configuration in \figref{fig:trans}(a). (b) One realization
  of obstacles randomly drawn from the 
corresponding probability distribution, with a non-overlapping
constraint.  
 Parameters
  used are: $\delta = 0.02$, $\epsilon = 0.01$.} 
\label{fig:random_s}
\end{figure}

To determine $\varepsilon({\bf x})$, 
consider one representative cell $ \mathcal A({\bf x})$
centered at ${\bf x}$ in the original domain $\Omega$.
The cell $\mathcal A ({\bf x})$ is mapped to a square of side
$\delta$ centered at ${\bf x}'$ in $\Omega'$, which we denote
$\mathcal A' ({\bf x}')$. 
The area of $\mathcal A ({\bf x})$ is then
\begin{equation}
|{\mathcal A} ({\bf x})| =\int_{\mathcal A ({\bf x}) } \ud {\bf x} = 
\int_{\mathcal A' ({\bf x}')}  \det J_W^{-1} \ud {\bf x}' =
\int_{\mathcal A'({\bf x}') } \frac{1}{\det J_W } \ud {\bf x}' =
\int_{\mathcal A' ({\bf x}')} \frac{\ud {\bf x}' }{ \| {\bf x}'
  \|^4}. \label{Aeqn}
\end{equation}
Since the volume fraction in the cell is conserved, we have 
\begin{equation}
\label{phi_exp}
\phi({\bf x}) = \frac{\pi \epsilon^2}{|{\mathcal A}({\bf x})|} =
\frac{\pi (\delta  \varepsilon({\bf x}))^2}{ \delta ^2} = \pi
\varepsilon({\bf x})^2. 
\end{equation}
Thus knowing $|{\mathcal A}({\bf x})|$ through \eqref{Aeqn} allows
us to determine both $\phi({\bf x})$ (and therefore $s({\bf x})$), and
also 
\begin{equation}
\label{vareps_exp}
\varepsilon({\bf x}) = \frac{\epsilon}{\sqrt{|{\mathcal A}({\bf x})|}},
\end{equation}
which we need in order to solve the
multiple-scales cell problem \eqref{cellproblem}.
In \figref{fig:random_s}(a) we plot the density $s({\bf x})$
corresponding to the configuration shown in \figref{fig:trans}(a). We
see that the maximum local solid volume fraction is about 2.5 times
the average solid volume fraction $\Phi$. In \figref{fig:random_s}(a)
we show one realization of obstacles randomly drawn from the
corresponding probability distribution, with a non-overlapping constraint.

We suppose that at $t=0$ a drop of solute of radius $\epsilon$ is
placed centered at ${\bf x}_0 = (0.0392, 0)$. We have 
chosen ${\bf x}_0$ such that the drop does not intersect with any of the obstacles, that is, $\| {\bf x}_0 -
{\bf r}_i \| > 2\epsilon$ for $i = 1, \dots, N_s$. Thus the initial
condition is $C({\bf x},0) =
1/(\pi \epsilon^2)$ for $\| {\bf x}- {\bf x}_0 \| \le \epsilon$, and zero
otherwise. Of course, this initial condition does not satisfy the
requirement of the multiple-scales method that it varies slowly with
respect to the obstacle separation; nevertheless we expect that it
will quickly spread into a function which does.

\begin{figure}[htb]
\begin{center}
\begin{overpic}[width=.24\linewidth,tics=10]{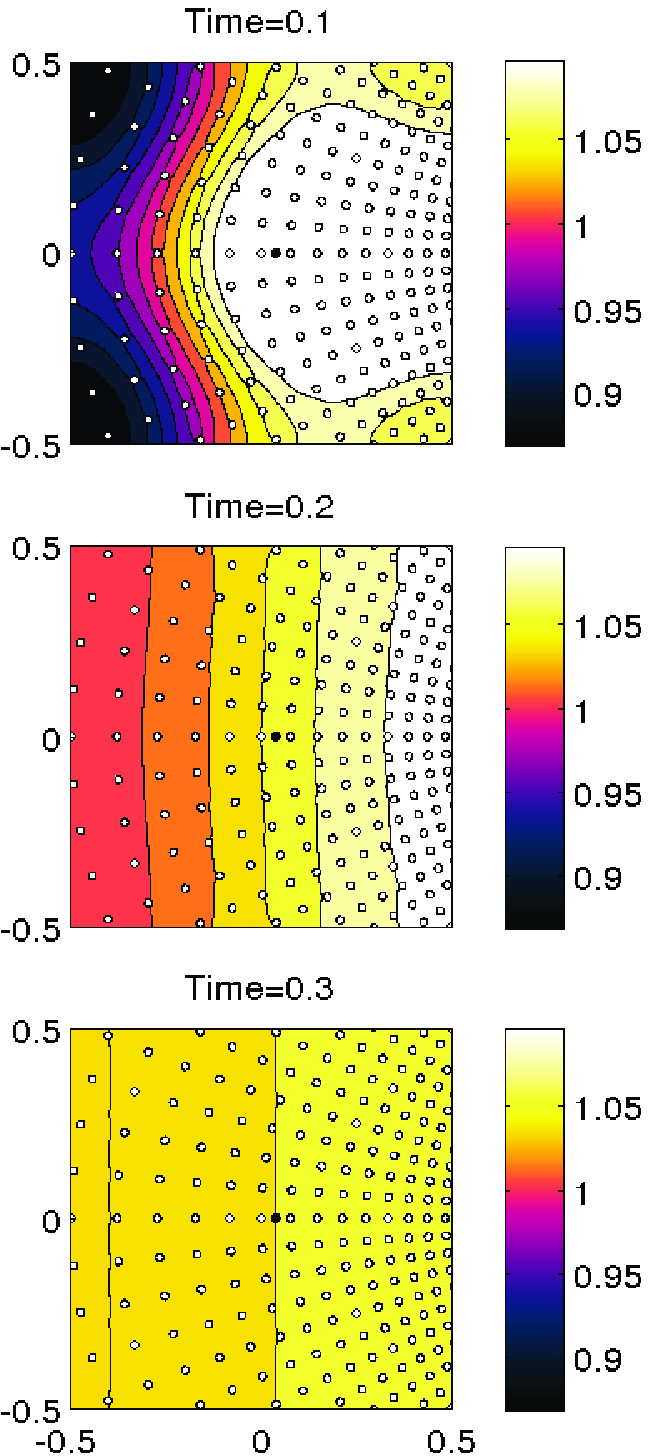} 
\put(0,98){\small (a)}
\end{overpic}
\begin{overpic}[width=.24\linewidth,tics=10]{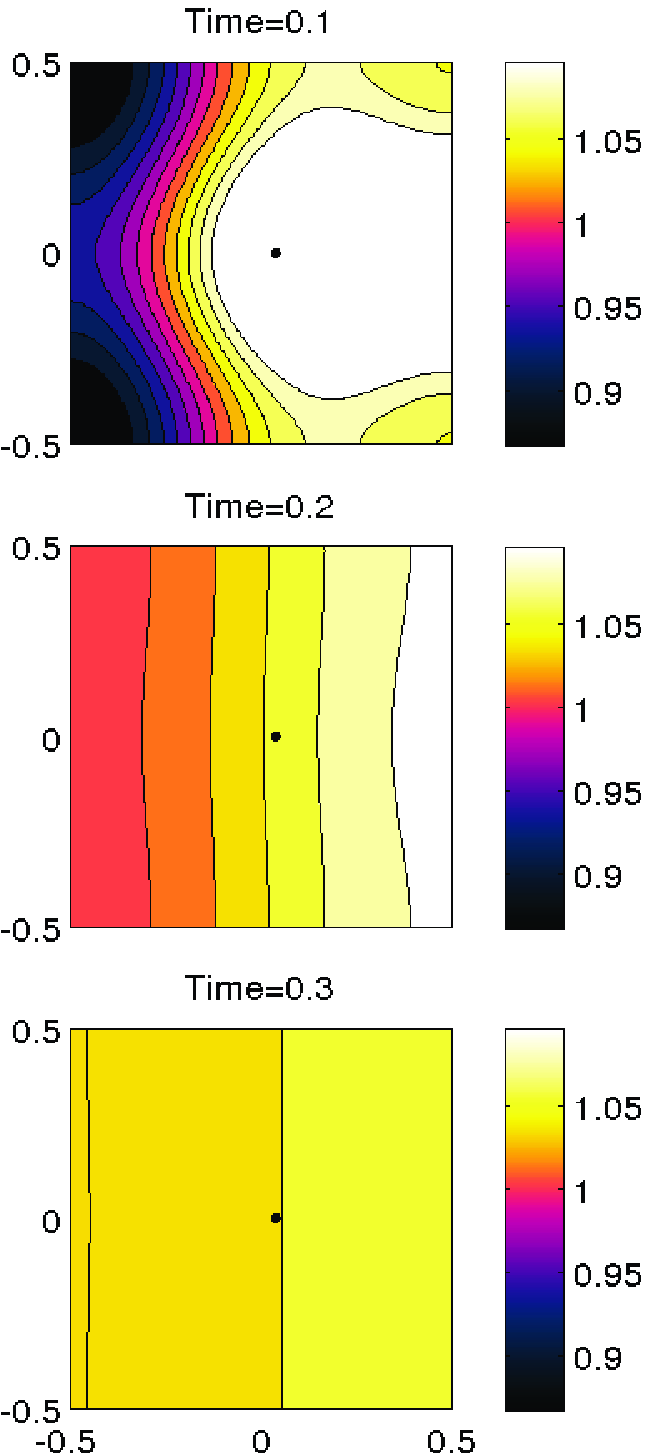} 
\put(0,98){\small (b)}
\end{overpic}
\begin{overpic}[width=.24\linewidth,tics=10]{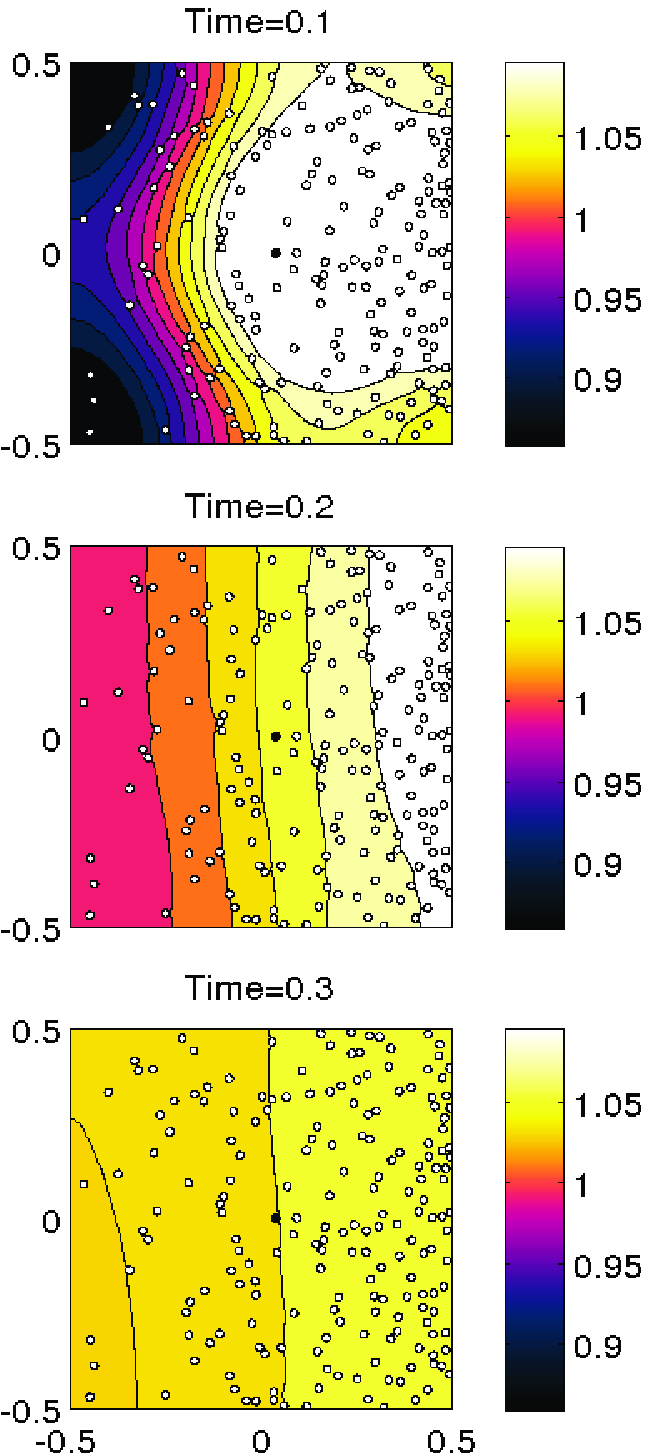} 
\put(0,98){\small (c)}
\end{overpic}
\begin{overpic}[width=.24\linewidth,tics=10]{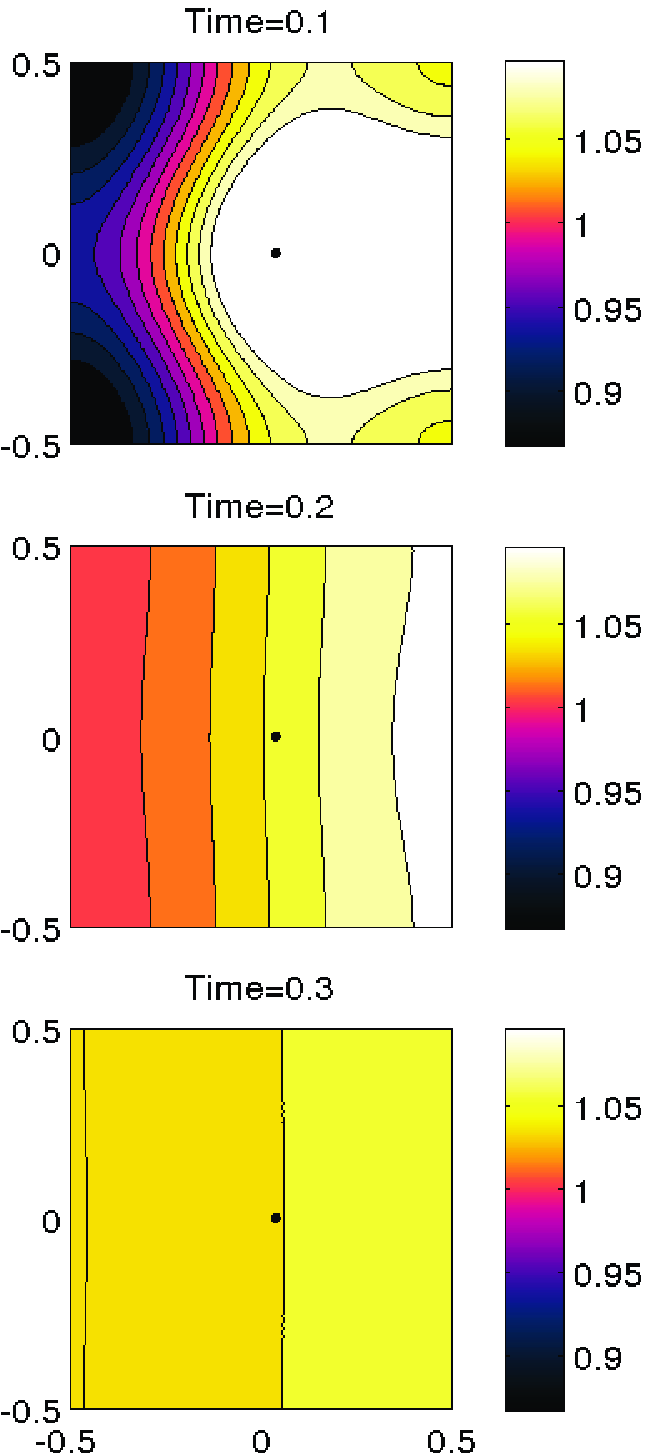} 
\put(0,98){\small (d)}
\end{overpic}
\end{center}
\caption{The concentration $C({\bf x},t)$ at times $t  = 0.1, 0.2, 0.3$ for  
(a) a locally periodic
distribution of obstacles; and (c) 
one realization of a
 random configuration of obstacles, distributed according to the
same  density function $s({\bf x})$. The intrinsic average $\bar{c}({\bf x},t)$ at the same times from  
(b) equation
\eqref{sol_poreav} derived through the
 multiple-scales method; and (d) the equation
 \eqref{asym_density} derived
 through the 
 Fokker--Planck approach. 
The position of the initial solute drop is shown
  as a black disk. } 
\label{fig:gradient_IA}
\end{figure}

In Figure \ref{fig:gradient_IA} 
we illustrate the time evolution of the solute density for four
different models. In  Figure~\ref{fig:gradient_IA}(a) we show the
(numerically calculated) true solution for the locally periodic
distribution of obstacles. In  Figure~\ref{fig:gradient_IA}(c) we show 
 the (numerically calculated) true solution for one realization of the
 random configuration of obstacles, distributed according to the
 density function $s({\bf x})$. In  Figure~\ref{fig:gradient_IA}(b) we
 show the solution of the effective equation  \eqref{sol_poreav}
 derived through the 
 multiple-scales method, while in
 Figure~\ref{fig:gradient_IA}(d) we 
 show the solution of the (intrinsic version of the) effective
 equation \eqref{asym_density} 
 derived through the 
 Fokker--Planck approach. In each case we show the 
intrinsic average $\bar c$, since this allows direct comparison
between the solutions of the effective equations and the solutions of
the real problems.\footnote{For the Fokker--Planck case, we recall that $p\equiv c$ given the normalization condition on $c$, and hence we identify the intrinsic average as $\bar c({\bf x},t) = c({\bf x},t)/\psi \equiv p({\bf x},t)/\psi$, where $\psi({\bf x}) = 1-\phi({\bf x})$.  }
 We see that, perhaps counterintuitively, the maximum of the solute
 concentration initially moves to the right, towards the region of low
 porosity. This is because the localized source spreads out in all
 directions, but spreads out more in the high porosity region. The
 increased diffusion in the high-porosity region lowers the
 concentration more, giving the impression that the solute is moving
 towards the low-porosity region.

At least visually, the two effective models capture the
behavior of the true solutions, both for the regular structure and
even for a single realization of the random structure.
To make the comparison a little easier, we show in
\figref{fig:cut_ui} a slice along the line $y=0$ for the true locally
periodic solution, the multiple-scales-derived solution, and the
Fokker--Planck-derived solution. We see that the agreement is
remarkably good.

\def \scc {0.7}
\def \scl {1.0}
\unitlength=.8cm
\begin{figure}[htb]
\centering
\psfragscanon
\psfrag{t}[l][][\scl]{$t$} 
\psfrag{x}[][][\scl]{$x$}
\psfrag{data1}[][][\scc]{MSpde}
\psfrag{data2}[][][\scc]{FPpde}
\psfrag{data3}[][][\scc]{MSfull}
\psfrag{ui}[][][\scl][-90]{$\bar c$}  
\includegraphics[width = .55\linewidth]{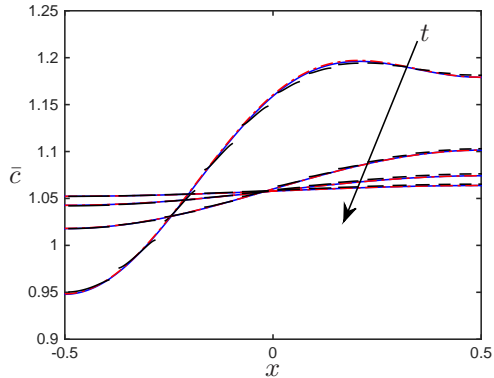} 
\caption{The intrinsic average $\bar c({\bf x},t)$ along
  the line $y=0$ at times $t = 0.1, 0.2, 0.3, 0.4$ computed from the multiple-scales equation
\eqref{sol_poreav}  (blue, full) and from the Fokker--Planck equation
 \eqref{asym_density} (red, dot-dashed). The true
  solution $C$ for a locally periodic 
distribution of obstacles is shown in black (broken by obstacles). }  
\label{fig:cut_ui}
\end{figure}

\begin{figure}[htb]
\begin{center}
\begin{overpic}[width=.24\linewidth,tics=10]{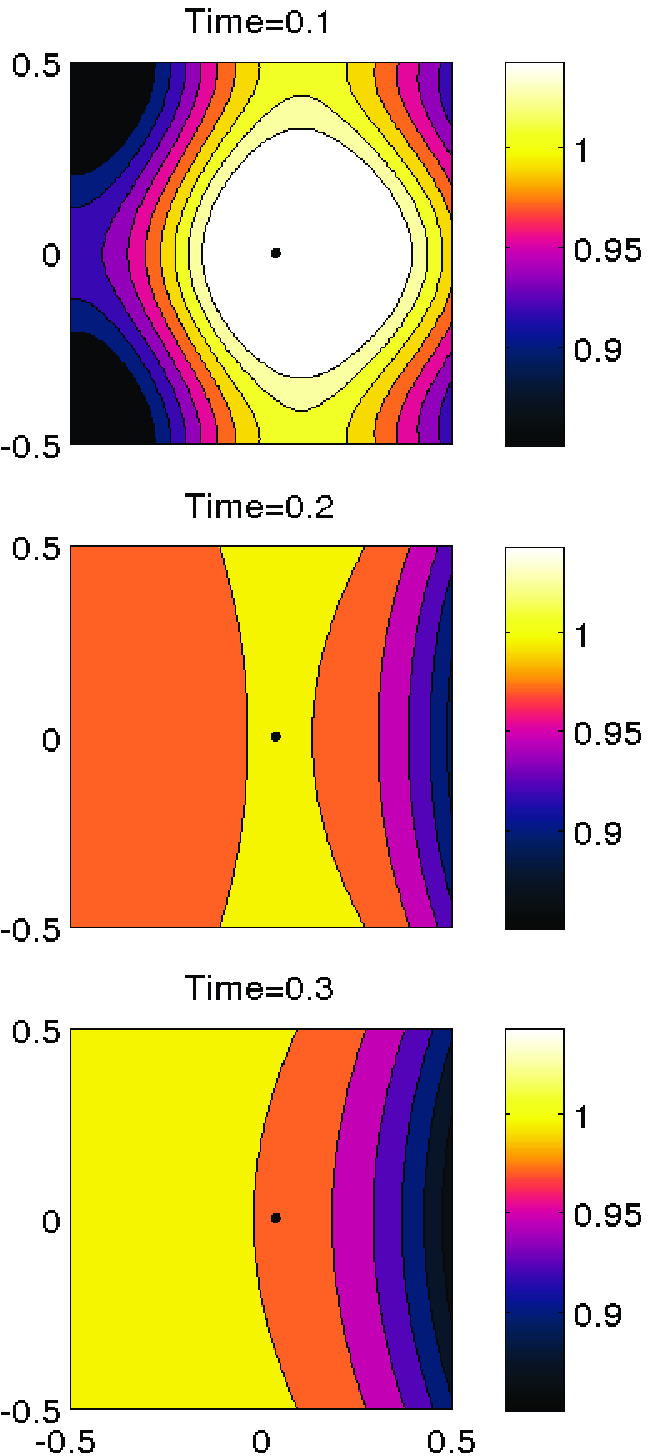} 
\put(0,98){\small (a)}
\end{overpic}
\quad
\begin{overpic}[width=.24\linewidth,tics=10]{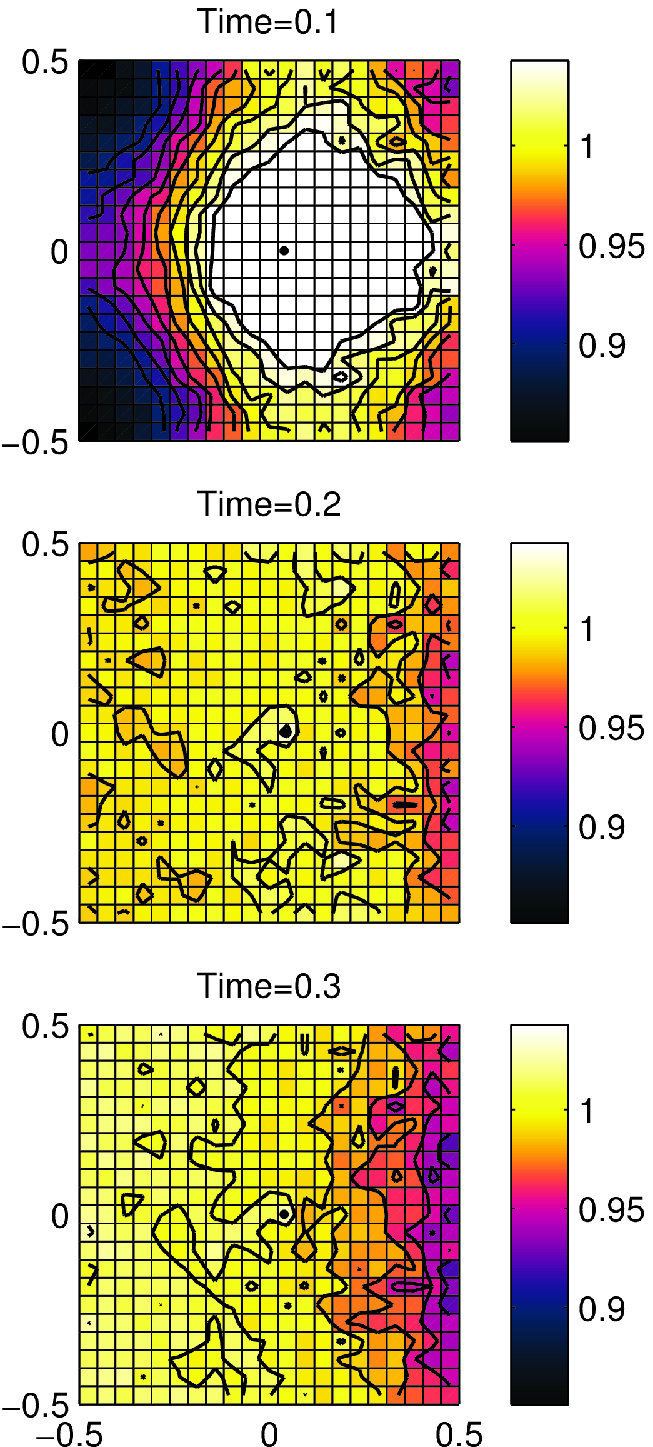} 
\put(0,98){\small (b)}
\end{overpic}
\quad
\begin{overpic}[width=.24\linewidth,tics=10]{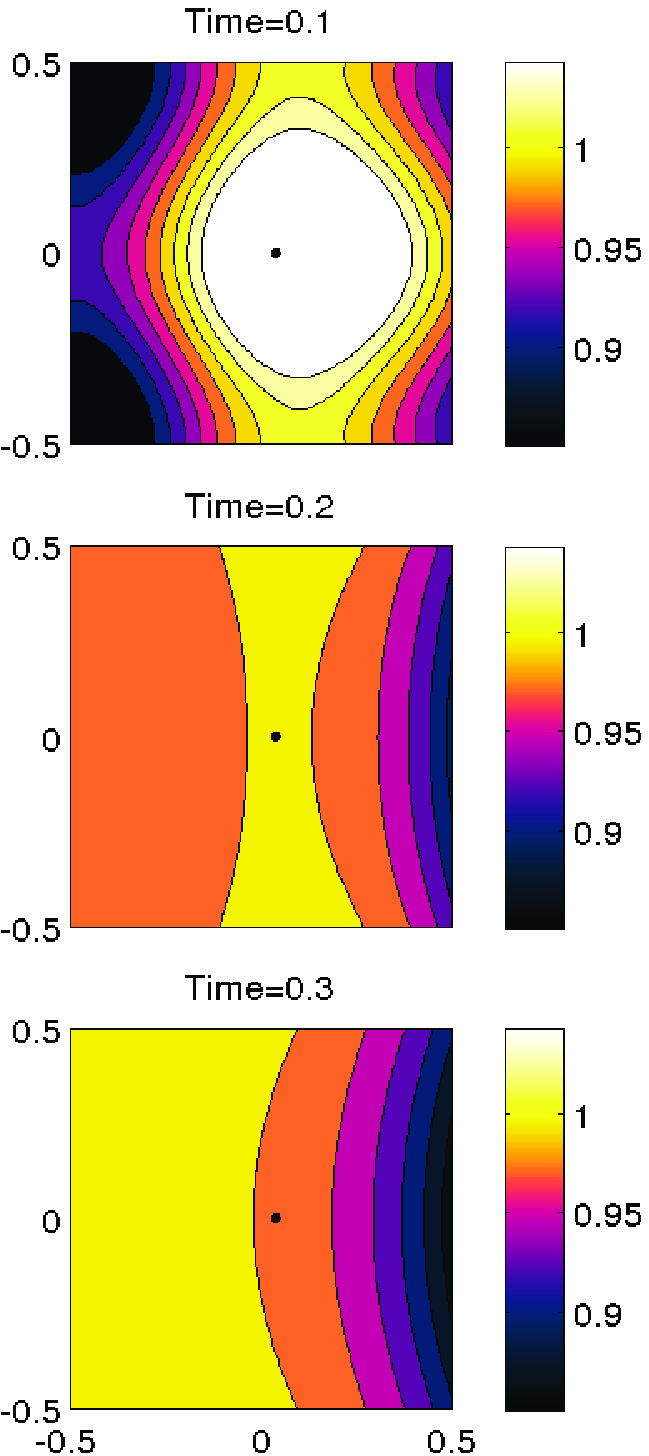} 
\put(0,98){\small (c)}
\end{overpic}
\end{center}
\caption{The volume average $c({\bf x},t)$ at times $t = 0.1, 0.2,
  0.3$ computed from (a) the effective equation
\eqref{sol_poreavmap} derived through the
 multiple-scales method; (b) 
the ensemble average of the true solution over 100 realizations of 
 random configurations of obstacles distributed according to the
 density function $s({\bf x})$; and (c) the effective equation
 \eqref{asym_density} derived
 through the 
 Fokker--Planck approach.
The position of the initial solute drop is shown
  as a black disk. 
} 
\label{fig:gradient_VA}
\end{figure}

While the intrinsic average $\bar c$ is the natural variable to compare with
a particular realization of the microstructure, the volume average $c$
is the natural variable to compare with an ensemble average over a
random microstructure (and is also usually the natural variable in an
application of the effective equations).
While $\bar c$ converges to the uniform
measure as time evolves (as in \figref{fig:gradient_IA}), the
 volume average $c$ approaches a non-uniform density which is a
 multiple of the porosity.
We show in \figref{fig:gradient_VA} the evolution of the  volume
average $c$ for three models.
In  Figure~\ref{fig:gradient_VA}(a) we show the
solution of the effective equation  \eqref{sol_poreavmap}
 derived through the 
 multiple-scales method, while in
 Figure~\ref{fig:gradient_VA}(c) we 
 show the solution of the effective equation \eqref{asym_density}
 derived through the 
 Fokker--Planck approach. In  Figure~\ref{fig:gradient_VA}(b) we show
 average of the true solution over 100 realizations of the randomly
 distributed obstacles.
Again, visually, both models seem to do a good job of approximating
the ensemble average.

In \figref{fig:cuty} we show the solutions in \figref{fig:gradient_VA} along the strip $y=[-3\Delta/2, 3\Delta/2]$ for $\Delta = 1/21$. In \figref{fig:cuty}(a) we show the solutions of the multiple-scales-derived and
Fokker--Planck-derived effective equations, along with the average over
100 realizations of the true solution for a random distribution of
obstacles (as in \figref{fig:gradient_VA}). In \figref{fig:cuty}(b)
we show some of the individual realizations along with the ensemble
average, to give an idea of the variance.

\def \scc {0.7}
\def \scl {.9}
\unitlength=.8cm
\begin{figure}[htb]
\centering
\psfragscanon
\psfrag{t}[b][][\scl]{$t$} 
\psfrag{a}[][][\scl]{(a)}
\psfrag{b}[][][\scl]{(b)}
\psfrag{x}[][][\scl]{$x$}
\psfrag{uv}[][][\scl][-90]{$c$}  
\psfrag{u}[][][\scl][-90]{$c$}  
\includegraphics[width = .45\linewidth]{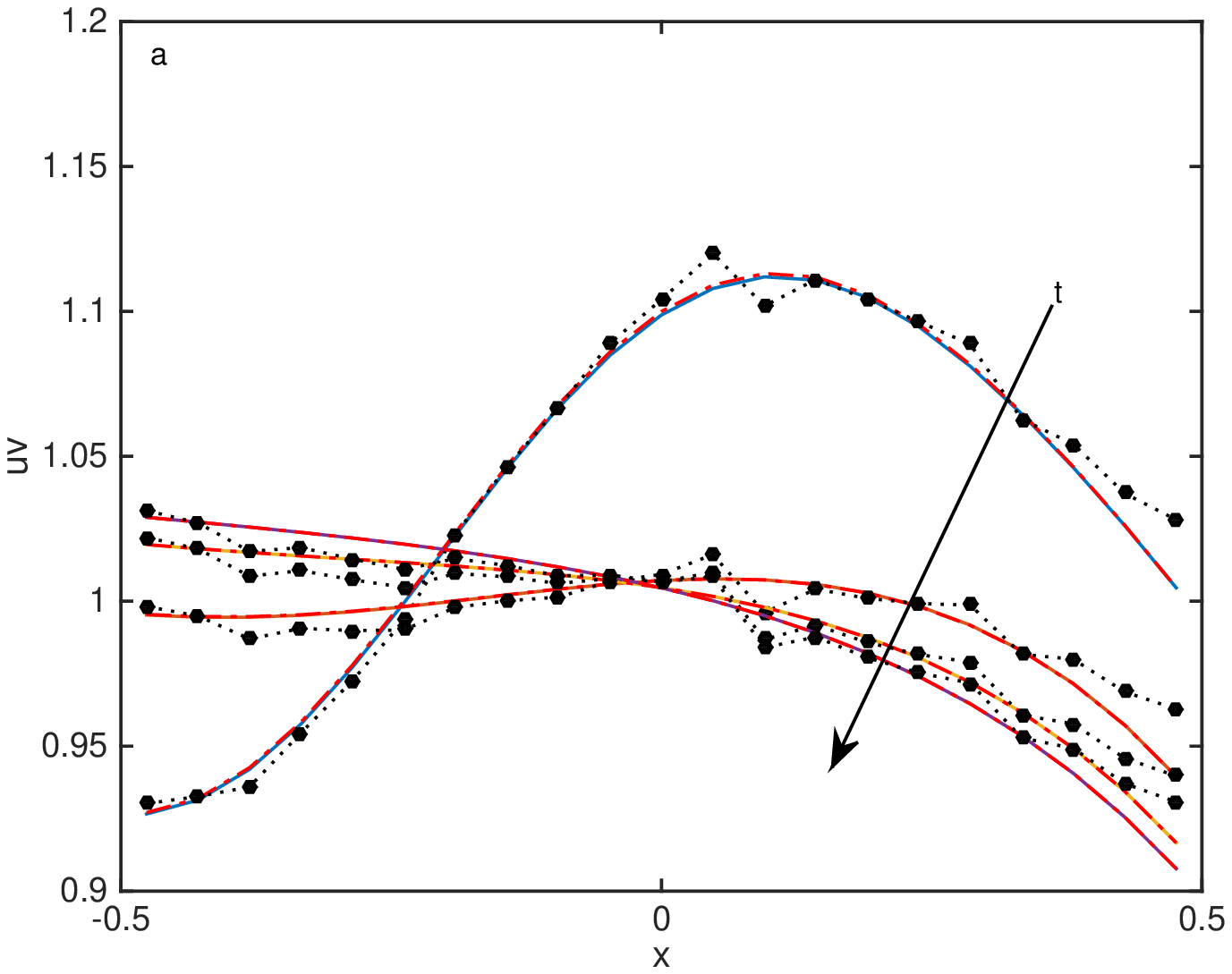} \quad
\includegraphics[width = .45\linewidth]{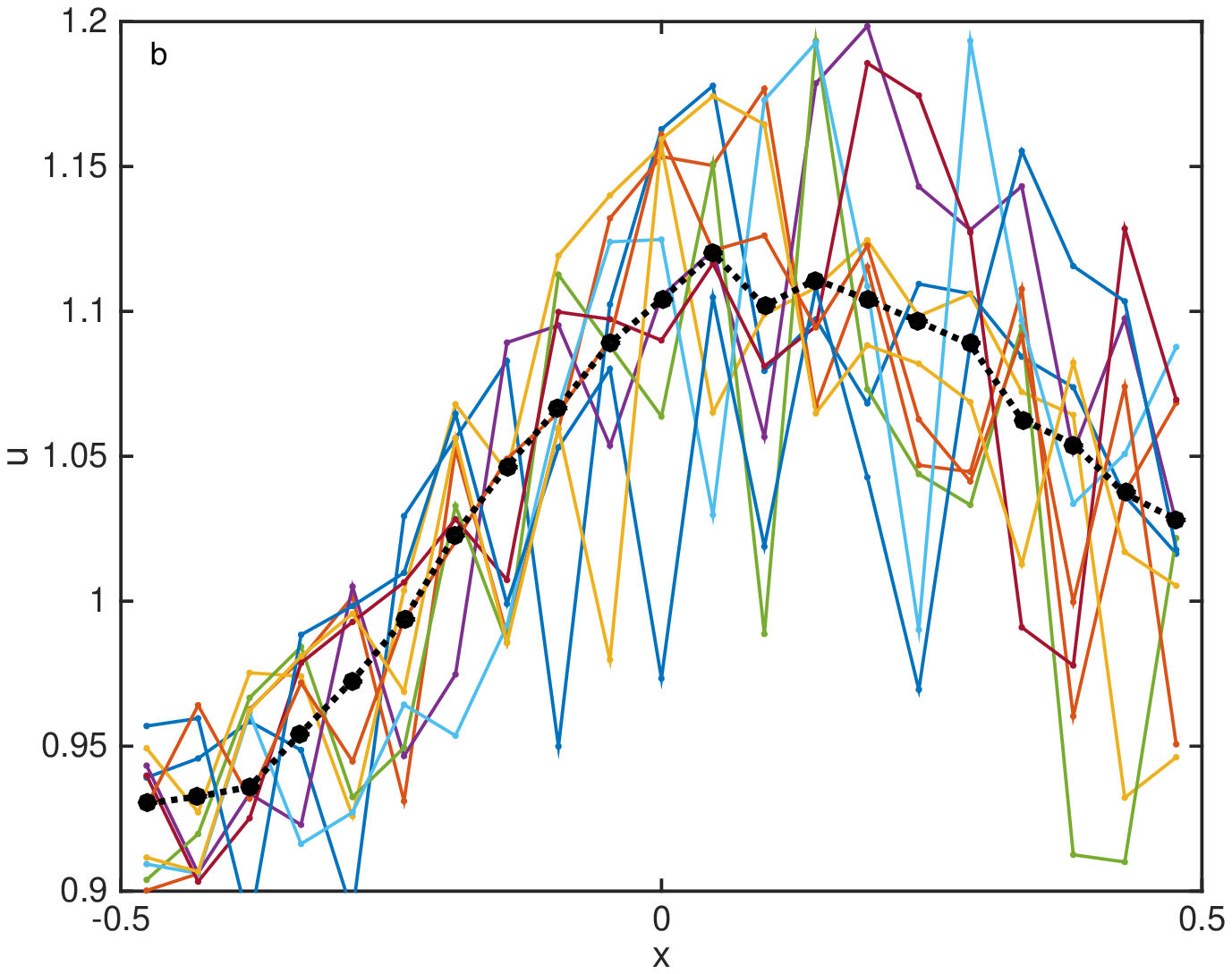} 
\caption{The volume average $c({\bf x},t)$  along the strip $y=[-3\Delta/2, 3\Delta /2]$, computed by averaging the solution over bins of width $\Delta = 1/21$ in $x$ and $3\Delta$ in $y$.
(a) Solution at times $t = 0.1, 0.2, 0.3, 0.4$ computed from the multiple-scales  equation \eqref{sol_poreav}  (blue, full) and the Fokker--Planck  equation \eqref{asym_density} (red, dot-dashed), and an ensemble average over 100 realizations of the true solution for a random distribution of obstacles (black dots). (b) Ensemble average (black dots) at time $t=0.1$ along with 10  individual realizations of the true solution (colored), again averaged over bins of width $\Delta=1/21$ in $x$ and $3\Delta$ in $y$.}
\label{fig:cuty}
\end{figure}

\section{Discussion} \label{sec:discussion}
 
We have investigated the problem of diffusion through a porous medium
in which the porosity is non-uniform.
Transport through porous media is inherently a multiscale process,
 with properties determined by  individual pores at the microscale
 while one is usually interested transport over much larger distances.
 To derive an effective
macroscopic model, a common approach is to assume that all
heterogeneities lie within the microscale and that, once these are averaged out,
we are left with a homogeneous porous medium at the macroscale. 
It is tempting to suppose that, if the properties such as porosity do
vary macroscopically, then all we have to do is carry out this
procedure locally at each point (that is, treat the material as though
it were uniform with a porosity equal to the local porosity), so that
the effective diffusivity becomes a function of macroscopic position. However, in this paper we have seen that
this is not the case, and that a more careful analysis is needed.

We have considered two different approaches to the upscaling problem, suitable for  deterministic and random porous media respectively, and have generalized them to heterogeneous media. The result is a macroscopic advection--diffusion equation, with the advection term accounting for the macroscopic gradients in porosity.  

First, we have extended the method of multiple scales to account for
non-uniform porosity (via a microscopic cell geometry parametrized by
the macroscopic variable) and non-periodic structures (providing they
are locally periodic, that is, they can be mapped into periodic
structures by a transformation depending only on the slow scale). The
resulting equation is equivalent at leading 
order to the one
presented in \cite{ValdesParada:2011dr} using a volume-averaging
approach, although some formally higher-order terms are
included in their unit cell problem (which leads them to some worrying
conclusions; for example, the effective diffusivity depends on the
location of the centroid in the unit cell). 
The multiple-scales method has the advantage of being  a systematic
asymptotic expansion  
 (for which higher-order terms could in principle be
calculated) which is able to handle any porosity.
Its disadvantage is that, even with our extensions,
 it requires the microstructure to be locally periodic.

Our examples have considered spherical obstacles that remained spherical when mapped to a periodic arrangement, which has greatly simplified parts of the presentation. However, the technique is more general. In particular
if the diffusion tensor is anisotropic (for example, if the inclusions
were ellipsoids instead of spheres) the technique demonstrates
systematically that the principal 
directions of the diffusion tensor would be aligned with the local
axes of the ellipsoids.

One question that always arises when deriving effective equations
with the method of multiple scales is how much of an error is
introduced by treating the microstructure as periodic, when in reality
it is unstructured.
To address this question, we considered a second approach
involving  diffusion 
through a random distribution of spherical obstacles. 
In this case we again used a systematic asymptotic expansion, but this
time in the limit of low volume fraction of obstacles.
The macroscopic equation
turns out to be a particular case of our model for the diffusion of
binary mixtures of finite-sized particles \cite{Bruna:2012wu}, when
the diffusivity of one of the species is set to zero.

We compared the macroscopic models described above to each other and
also to their microscopic counterparts. For the examples we considered both effective
material models performed 
well, and the differences between the structured and unstructured
media were small. 

For both deterministic and random media, the mean-square displacement $\langle r^2 \rangle$ analysis in \figref{fig:msd_simul} showed that the effective transport mechanism of the solute particle is still normal diffusion, that is, $\langle r^2 \rangle$ grows linearly in time, and it is only the diffusion coefficient that changes from one (in the microscale) to $D_e$ in the macroscale. However, depending on the timescale used in such analysis, one can observe \emph{transient} anomalous diffusion (the transient period can only just be seen in our simulations, see Figure \ref{fig:msd_simul}(b)). In the context of porous media, if the ratio $\epsilon$ between pore scale and typical domain length is not sufficiently small, diffusion may occur in a lengthscale shorter than the crossover region between transient anomalous diffusion and normal diffusion \cite{Saxton:1994hk}. This transient behavior should not be confused with real \emph{stationary} anomalous diffusion \cite{Berezhkovskii:2014gb}. 

\Appendix

\section{Numerical convergence of mean-square displacement simulations} \label{sec:convergence}
Using the method described in \S\ref{sec:self-diffusion}, we compute the effective diffusion coefficient in porous media with 20\% obstacle volume fraction, with obstacles either on a square lattice or uniformly distributed. We use the same parameters as in \figref{fig:msd_simul} except the timestep $\Delta t$, which we will vary to perform a convergence study. In addition, we use this study to check the number of trajectories (the number of particles $N_m$ times the number of runs $M$) required to obtain an accurate mean-square displacement. 
This is particularly important in the random case, as we need to obtain an accurate mean from the distribution of random media. 

The mean displacement of a diffusive particle according to  \eqref{sde_b} is $h =\sqrt{2 \Delta t}$. If the obstacle's diameter is $2 \epsilon$, then we should use $\Delta t $ such that $h \ll 2\epsilon$. We use a simulation time $t=0.25$  (so that each particle has had time to diffuse across almost the whole domain $\Omega$). We choose a mean step $h = (2\epsilon)/2^k$ for $k = 0, 1, \dots, 4$, or, equivalently, a timestep $\Delta t = \epsilon^2/2^{2k-1}$, where $\epsilon = 0.0126$. The diffusion coefficients are obtained as the average of the last 10 data points of the time series (after checking it has converged to a stationary value, see dashed area in \figref{fig:msd_simul}(b)). The resulting values for the periodic and random media, denoted by $D_e^{(1)}$ and $\hat D_e^{(1)}$ respectively, are shown in  Table \ref{table:De}. The standard deviation in both cases is below 0.006 for all $k$.

\begin{table}[htbp]
\caption{Diffusion coefficient in a square lattice with volume fraction $\Phi = 20\%$ from simulations using $\Delta t = \epsilon^2/2^{2k-1}$. Theoretical predictions $D_e$ and $\hat D_e$ from Eqs.  \eqref{Dxm} and \eqref{point_Dcollective} respectively are also shown.}
\begin{center}\footnotesize
\renewcommand{\arraystretch}{1.3}
\begin{tabular}{| l | l | l | l | l | l | l | }\hline
& $D_e^{(1)}$  & $D_e^{(2)}$ & $D_e^{(3)}$ & $D_e^{(4)}$ & $D_e^{(5)}$ & $D_e$ \\  \hline
$k = 0$ & 0.984483  & 0.860079 & 0.790490& 0.833896 & 0.833425 & 0.833163\\ 
$k = 1$ & 0.891180 & 0.794840& 0.833218 & 0.833427 & &\\ 
$k = 2$ & 0.818925 & 0.830819 & 0.833424 &  & &\\ 
$k = 3$ & 0.827845 & 0.833261 & &  & & \\ 
$k = 4$ & 0.831907 & & &  & & \\ \hline   \hline
& $\hat D_e^{(1)}$  & $\hat D_e^{(2)}$ & $\hat D_e^{(3)}$ & $\hat D_e^{(4)}$ & $\hat D_e^{(5)}$ & $\hat D_e$ \\ \hline
$k = 0$ &  0.895929	& 0.815169 & 0.779048 &	0.811383 & 0.807609 & 0.8\\ 
$k = 1$ & 0.835359 & 0.781305 & 0.810877 & 0.807624 & &\\ 
$k = 2$ & 0.794819 & 0.809029 & 0.807675 &  & &\\ 
$k = 3$ & 0.805476 & 0.807759 & &  & &\\ 
$k = 4$ & 0.810293 & & &  & & \\ \hline   
\end{tabular}
\end{center}
\label{table:De}
\end{table}

The order of convergence of $D_e^{(1)}$ and $\hat D_e^{(1)}$ as $\Delta t\to 0$ is one (in $\Delta t$), as expected from the Euler--Maruyama integration scheme. We can apply repeated Richardson extrapolations $D_e^{(i)}$  and $\hat D_e^{(i)}$ to improve the accuracy of these numerical results (see Table \ref{table:De}). By the last extrapolation,  $D_e^{(5)}$ appears to have five correct figures, three of which coincide with the theoretical value $D_e$. Similarly, $\hat D_e^{(5)}$ has four correct figures, two of which agree with the theoretical value $\hat D_e$. 
\newline

\textbf{Acknowledgements.} The authors thank Dr Martin Robinson for helpful discussions on the implementation of the stochastic algorithms. In particular, we benefited from his software library Aboria (available online at https:$/\!/$github.com/martinjrobins/ Aboria), which includes the neighbourhood search subroutine we have used to speed up the calculation of collisions in the stochastic simulations.

\end{document}